\documentclass[12pt,bibliography=totoc,DIV=17]{scrartcl}
\pdfoutput=1
\usepackage[utf8]{inputenc}

\usepackage{amsmath,amsfonts,amssymb,dsfont,mathtools}
\usepackage{array}
\usepackage[symbol]{footmisc}
\usepackage{stackrel}
\let\refeq\undefined
\usepackage[mathscr]{euscript}
\usepackage{enumitem}
\usepackage{longtable}
\usepackage{booktabs}
\usepackage{multirow}
\usepackage{cite}
\usepackage{graphicx, ulem, hhline}
\usepackage[table,dvipsnames]{xcolor}
\usepackage{xspace}
\usepackage{needspace}
\usepackage{siunitx}
\usepackage{soul}
\usepackage{wasysym}              
\usepackage{hyphenat}             

\usepackage{afterpage}
\usepackage{etoolbox,xstring,xspace,calc,xifthen}
\usepackage{centernot}

\usepackage{tikz}
\usetikzlibrary{tikzmark, calc}

\tikzset{
    double color fill/.code 2 args={
        \pgfdeclareverticalshading[%
            tikz@axis@top,tikz@axis@middle,tikz@axis@bottom%
        ]{diagonalfill}{100bp}{%
            color(0bp)=(tikz@axis@bottom);
            color(50bp)=(tikz@axis@bottom);
            color(50bp)=(tikz@axis@middle);
            color(50bp)=(tikz@axis@top);
            color(100bp)=(tikz@axis@top)
        }
        \tikzset{shade, left color=#1, right color=#2, shading=diagonalfill}
    }
}
\tikzset{
    stripes fill/.code 2 args={
        \pgfdeclareverticalshading[%
            tikz@axis@top,tikz@axis@middle,tikz@axis@bottom%
        ]{diagonalfill}{100bp}{%
            color(0bp)=(tikz@axis@bottom);
            color(12.5bp)=(tikz@axis@bottom);
            color(25bp)=(tikz@axis@bottom);
            color(25bp)=(tikz@axis@top);
            color(37.5bp)=(tikz@axis@top);
            color(37.5bp)=(tikz@axis@bottom!99);
            color(50bp)=(tikz@axis@bottom!99);
            color(50bp)=(tikz@axis@top!99);
            color(62.5bp)=(tikz@axis@top!99);
            color(62.5bp)=(tikz@axis@bottom);
            color(75bp)=(tikz@axis@bottom);
            color(75bp)=(tikz@axis@top);
            color(100bp)=(tikz@axis@top)
        }
        \tikzset{shade, left color=#1, right color=#2, shading=diagonalfill}
    }
}
\newcounter{BGnum}
\makeatletter
\newif\iftag@here
\newcommand*{\taghere}[1][0pt]
{\ifmeasuring@\else
  \global\tag@heretrue
  \tikz[remember picture,overlay]{\coordinate (taghere) at (0pt,#1);}%
\fi}

\def\place@tag{%
    \iftagsleft@
      \kern-\tagshift@
      \iftag@here
        \global\tag@herefalse
        \tikz[remember picture,overlay]%
          {\path (taghere) -| node[anchor=base]{\rlap{\boxz@}} (0pt,0pt);}%
      \else
        \if1\shift@tag\row@\relax
            \rlap{\vbox{%
                \normalbaselines
                \boxz@
                \vbox to\lineht@{}%
                \raise@tag
            }}%
        \else
            \rlap{\boxz@}%
        \fi
        \kern\displaywidth@
      \fi
    \else
      \kern-\tagshift@
      \iftag@here
        \global\tag@herefalse
        \tikz[remember picture,overlay]%
          {\path  (taghere) -|  node[anchor=base]{\llap{\boxz@}} (0pt,0pt);}%
      \else
        \if1\shift@tag\row@\relax
            \llap{\vtop{%
                \raise@tag
                \normalbaselines
                \setbox\@ne\null
                \dp\@ne\lineht@
                \box\@ne
                \boxz@
            }}%
        \else \llap{\boxz@}%
        \fi
      \fi
    \fi
}
\makeatother

\def\thefootnote{\arabic{footnote}}

\makeatletter
\newlength{\fnhskip}
\renewcommand\@makefntext[1]{
  \settowidth{\fnhskip}{\@makefnmark}
  \leftskip=\fnhskip
  \hskip-\fnhskip
  \@makefnmark#1
}
\makeatother

\renewenvironment{subequations}[1][]{
  \refstepcounter{equation}%
  \setcounter{parentequation}{\value{equation}}
  \setcounter{equation}{0}
  \def\theequation{\theparentequation\alph{equation}}%
  \let\parentlabel\label
  \ifx\\#1\\\relax\else\label{#1}\fi
  \ignorespaces
}{%
  \setcounter{equation}{\value{parentequation}}
  \ignorespacesafterend
}

\newcommand*{\nextParentEquation}[1][]{
  \refstepcounter{parentequation}
  \setcounter{equation}{0}
  \ifx\\#1\\\relax\else\parentlabel{#1}\fi
}

\usepackage{setspace}

\usepackage[numbers,sort&compress]{natbib}
\makeatletter
\def\NAT@spacechar{\,}
\makeatother
\usepackage[
colorlinks=true
,urlcolor=blue
,anchorcolor=blue
,citecolor=blue
,filecolor=blue
,linkcolor=blue
,menucolor=blue
,linktoc=all
,unicode
,backref=page
]{hyperref}
\usepackage{hypernat}

\newrobustcmd*{\tocref}[1]{\hyperref[TOC]{\color{black}{#1}}}
\usepackage[all]{hypcap}
\usepackage{footnotebackref}

\usepackage[format=plain, margin=0.5cm, font=footnotesize, labelfont=bf, list=off]{caption}

\renewcommand*{\backref}[1]{}
\renewcommand*{\backrefalt}[4]{%
  \ifcase #1%
  \or [p\,#2]%
  \else [pp\,#2]%
  \fi%
}

\newif\ifbackrefshowonlyfirst
\backrefshowonlyfirsttrue
\makeatletter
\let\BR@direct@old@hyper@natlinkstart\hyper@natlinkstart
\renewcommand*{\hyper@natlinkstart}{\phantomsection\BR@direct@old@hyper@natlinkstart}
\let\BR@direct@oldBR@citex\BR@citex
\renewcommand*{\BR@citex}{\phantomsection\BR@direct@oldBR@citex}%

\long\def\hyper@page@BR@direct@ref#1#2#3{\hyperlink{#3}{#1}}

\ifx\backrefxxx\hyper@page@backref
    \let\backrefxxx\hyper@page@BR@direct@ref
    \ifbackrefshowonlyfirst
    \fi
\else
    \ifbackrefshowonlyfirst
    \fi
\fi

\patchcmd{\Hy@backout}{Doc-Start}{\@currentHref}{}{\errmessage{I can't seem to patch backref}}
\makeatother

\let\theparentequation\theequation
\patchcmd{\theparentequation}{equation}{parentequation}{}{}

\apptocmd{\thebibliography}{\scriptsize}{}{}
\let\OLDthebibliography\thebibliography
\renewcommand\thebibliography[1]{
  \OLDthebibliography{#1}
  \setlength{\parskip}{1pt}
  \setlength{\itemsep}{1pt plus 0.3ex}
}

\addtokomafont{disposition}{\rmfamily}
\BeforeTOCHead[toc]{{\pdfbookmark[1]{\contentsname}{toc}}}

\makeatletter
\patchcmd{\upbracefill}{\m@th}{\scriptstyle\m@th}{}{}
\patchcmd{\upbracefill}{$\braceld$}{$\scriptstyle\braceld$}{}{}
\patchcmd{\upbracefill}{\bracelu}{\bracelu\mkern-1mu}{}{}
\patchcmd{\upbracefill}{\hfill\braceru}{\hfill\mkern-1mu\braceru}{}{}
\DeclareOldFontCommand{\rm}{\normalfont\rmfamily}{\mathrm}
\DeclareOldFontCommand{\sf}{\normalfont\sffamily}{\mathsf}
\DeclareOldFontCommand{\tt}{\normalfont\ttfamily}{\mathtt}
\DeclareOldFontCommand{\bf}{\normalfont\bfseries}{\mathbf}
\DeclareOldFontCommand{\it}{\normalfont\itshape}{\mathit}
\makeatother

\newlength{\floatwidth}
\def\beq{\begin{equation}}
\def\eeq{\end{equation}}

\newcommand{\Amp}[4][\mathcal{A}]{#1^{\mbox{\tiny #2}}_{\mbox{\tiny #3}}\ifthenelse{\isempty{#4}}{}{{\left[#4\right]}}}

\def\twomat[#1,#2][#3,#4]{\left( \begin{array}{cc} #1 & #2 \\ #3 & #4 \end{array} \right)}
\def\threemat[#1,#2,#3][#4,#5,#6][#7,#8,#9]{\left( \begin{array}{ccc} #1 & #2 & #3\\ #4 & #5 & #6 \\ #7 & #8 & #9 \end{array} \right)}
\def\twovec[#1,#2]{\left( \begin{array}{c} #1  \\ #2 \end{array} \right)}
\def\thv[#1,#2,#3]{\left( \begin{array}{c} #1 \\ #2 \\ #3 \end{array} \right)}
\def\twv[#1,#2]{\left( \begin{array}{c} #1 \\ #2 \end{array} \right)}

\newcommand{\IE}{\textit{i.\,e.}\xspace}
\newcommand{\EG}{\textit{e.\,g.}\xspace}
\newcommand{\AP}{\mbox{\textit{a~priori}}\xspace}

\newcommand{\refeq}[1]{Eq.\,\eqref{#1}}
\newcommand{\refeqs}[1]{Eqs.\,\eqref{#1}}
\newcommand{\sect}[1]{Sect.\,\ref{#1}}
\newcommand{\appx}[1]{Appx.\,\ref{#1}}
\newcommand{\fig}[1]{Fig.\,\ref{#1}}
\newcommand{\tab}[1]{Tab.\,\ref{#1}}
\newcommand{\citere}[1]{Ref.\,\cite{#1}}
\newcommand{\citeres}[1]{Refs.\,\cite{#1}}
\newcommand{\simord}{\mathord{\sim}\,}

\newcommand{\gsim}{\gtrsim}

\newcommand{\yint}[3]{Y_{#1}^{[#2]}\ifthenelse{\isempty{#3}}{}{{\left(#3\right)}}}
\newcommand{\ysum}[3]{Y_{#1}^{#2}\ifthenelse{\isempty{#3}}{}{{\left(#3\right)}}}
\newcommand{\yonesum}[3]{{^0}Y_{#1}^{#2}\ifthenelse{\isempty{#3}}{}{{\left(#3\right)}}}
\newcommand{\tint}[2]{\ifthenelse{\isempty{#2}}{T_{#1}}{T_{#1}^{\left|#2\right.}}}

\newcommand{\toneint}[2]{\ifthenelse{\isempty{#2}}{{^0}T_{#1}}{^0T^{\left|#2\right.}_{#1}}}

\newcommand{\yoneint}[3]{{^0}Y_{#1}^{[#2]}\ifthenelse{\isempty{#3}}{}{{\left(#3\right)}}}
\newcommand{\bint}[1]{B_{0}\ifthenelse{\isempty{#1}}{}{{\left(#1\right)}}}
\newcommand{\aint}[1]{A_{0}\ifthenelse{\isempty{#1}}{}{{\left(#1\right)}}}
\newcommand{\binteps}[2]{\ifthenelse{\isempty{#2}}{B_{0}^{\left|#1\right.}}{B_{0}^{\left|#1\right.}{\!\left(#2\right)}}}

\newcommand{\CP}{\ensuremath{\mathcal{CP}}\xspace}
\newcommand{\MS}{\ensuremath{\overline{\text{MS}}}\xspace}
\newcommand{\DR}{\ensuremath{\overline{\text{DR}}}\xspace}

\newcounter{notecount}
\hypersetup{ pdfauthor={Me, myself and I}
	,pdftitle={About the bosonic decays of heavy Higgs states in the (N)MSSM} ,pdfsubject={}
	,pdfkeywords={} }

\begin{document}
\newcommand*{\mytitle}[1]{%
  \parbox{\linewidth}{\setstretch{1.5}\centering\Large\textsc{\textbf{\boldmath #1}}}
}

\thispagestyle{empty}

\def\thefootnote{\fnsymbol{footnote}}

\begin{flushright}
  BONN-TH-2022-16\\
  TTK-22-23
\end{flushright}

\vfill

\begin{center}

\mytitle{About the bosonic decays of\\ heavy Higgs states in the (N)MSSM}

\vspace{1cm}

Florian Domingo$^{1}$\footnote{email: domingo@physik.uni-bonn.de}
and
Sebastian Pa{\ss}ehr$^{2}$\footnote{email: passehr@physik.rwth-aachen.de}

\vspace*{1cm}

\textsl{
$^1$Bethe Center for Theoretical Physics \&
Physikalisches Institut der Universit\"at Bonn,\\
Nu\ss allee 12, D--53115 Bonn, Germany
}

\medskip
$^2$\textsl{Institute for Theoretical Particle Physics and Cosmology,}\\
\textsl{RWTH Aachen University, Sommerfeldstra{\ss}e 16, 52074 Aachen, Germany.}

\end{center}

\vfill

\begin{abstract}{}
The heavy, doublet-dominated Higgs bosons expected in
Two-Higgs-Doublet-Model-like extensions of the Standard Model are
obvious targets for searches at high-energy colliders, and
considerable activity in this sense is currently employed in analyzing
the results of the Large Hadron Collider. For sufficiently heavy
states, the $SU(2)_\mathrm{L}$ symmetry drastically constrains the
decays of these new scalars. In models that contain only additional Higgs doublets, fermionic
decay channels are expected to dominate, although some bosonic modes
may lead to cleaner signals. The situation is very different if
the Higgs sector is further extended by singlet states and
  the singlet-dominated scalars are kinematically accessible, in
which case the Higgs-to-Higgs widths may even become dominant. Yet, a
quantitative interpretation of the experimental data in terms of a specific model also runs through the
control of radiative corrections in this model. In this paper, we
examine the bosonic decays of the heavy, doublet-dominated Higgs
states in the~MSSM and the~NMSSM at full one-loop order, insisting on
the impact of the $SU(2)_\mathrm{L}$ symmetry and the effects of
infrared type, in particular, the emergence of sizable non-resonant
contributions to the three-boson final states.
\end{abstract}

\vfill
\def\thefootnote{\arabic{footnote}}
\setcounter{page}{0}
\setcounter{footnote}{0}
\newpage
\section{Introduction}

One of the possible forms of physics beyond the
Standard Model~(SM) consists in an extended
Higgs sector, with new states beyond the SM-like one, which has been
observed at the Large~Hadron~Collider~(LHC) with a mass of
  about~$125.25$\,GeV\,\cite{Aad:2015zhl,CMS:2020xrn,ATLAS:2018tdk}.
Considering the consequences of $SU(2)_\mathrm{L}$ triplet scalars (or
electroweak representations of higher dimensions) for the electroweak
symmetry breaking and the properties of the gauge bosons, the most
popular models essentially involve additional scalar doublets and
singlets, such as the Two-Higgs Doublet
Model~(THDM)\,\cite{Branco:2011iw}. Interestingly,
Supersymmetry~(SUSY), with its motivations in terms of a stabilization
of the electroweak scale, immediately implies a two-doublet structure
for phenomenological consistency in its most economical extension of
the~Standard~Model~(SM),
the~Minimal~Supersymmetric~SM~(MSSM)\,\cite{Nilles:1983ge,Haber:1984rc}.
Further singlet degrees of freedom are present in
the~Next-to-MSSM~(NMSSM)\,\cite{Maniatis:2009re,Ellwanger:2009dp},
where a singlet superfield is introduced as a solution to the
$\mu$-problem of the~MSSM\,\cite{Kim:1983dt}.

The measured properties of the SM-like
Higgs\,\cite{CMS-HL,Sirunyan:2018koj,Aad:2019mbh}, flavor
physics\,\cite{Misiak:2017zan} or direct
searches\,\cite{CMS:2018rmh,CMS:2018amk,CMS:2019mij,CMS:2019pzc,Aad:2020zxo,ATLAS:2020tlo,ATLAS:2021uiz}
corroborate the phenomenological necessity for new doublet-dominated
Higgs bosons, if they exist, to be comparatively heavy, in which case
an effective~SM emerges at low energy. Then, in this decoupling limit
with~$M_{H^{\pm}}\gg M_Z$, the global electroweak symmetry controls
the high-energy phenomenology and the \CP-even~$H$, \CP-odd~$A$ and
charged Higgs states~$H^{\pm}$ form an almost degenerate
$SU(2)_\mathrm{L}$~doublet. Provided they are kinematically
accessible, such doublet-dominated states could be directly produced
at the~LHC with the conventional mechanisms (gluon--gluon fusion,
assisted production, etc.). Due to the approximate
$SU(2)_\mathrm{L}$~symmetry, their decay widths are expected to be
controlled by the fermionic modes, and the latter---in particular the
cleaner leptonic final states---count among the active search channels
at the~LHC, see
\EG~\citeres{CMS:2018rmh,CMS:2019mij,CMS:2019pzc,Aad:2020zxo}. On the
contrary, decays of the heavy doublet-Higgs states into electroweak
gauge or SM-like Higgs bosons usually depend on electroweak-symmetry
breaking effects, resulting in subleading rates at energies much
larger than the $SU(2)_\mathrm{L}$-breaking scale; nevertheless such
channels may provide clean signals at colliders.

The situation changes drastically in the presence of singlet-dominated
states (in addition to the heavy doublet-like Higgs bosons). The
latter are typically difficult to directly produce and test in
(\EG)~proton collisions, due to their suppressed couplings to
SM~fermions and gauge bosons---unless an accidentally large
singlet--doublet mixing occurs. On the other hand, the singlet states
may induce cascade decays of the more-easily produced doublet
states---if kinematically accessible---and considerably complicate the
visibility of the latter if such Higgs-to-Higgs modes are dominant. We
refer the reader to the recent \citere{Ellwanger:2022jtd}, and
references therein, for search proposals exploiting Higgs-to-Higgs
transitions at the~LHC. Experimental limits considering such cascade
decays\,\cite{CMS:2019qcx,CMS:2019kca,CMS:2021klu,CMS:2021xor,CMS:2019ogx,CMS:2022qww,ATLAS:2020gxx,ATLAS:2022enb}
have also appeared in the last few years, demonstrating the potential
of the~LHC for investigating such scenarios.

In this paper, we focus on the theoretical aspects of the bosonic
decays of heavy, doublet-dominated Higgs states in
the~(N)MSSM. Indeed, the interpretation of experimental constraints in
a particular model crucially depends on the degree of control achieved
in the theoretical predictions of this model, in particular for the
decay widths and branching ratios. We aim at a description of Higgs
decays at the full one-loop~(1L) order (and higher orders in~QCD), as
already underlined in earlier stages of this
project\,\cite{Domingo:2017rhb,Domingo:2018uim,Domingo:2019vit,Domingo:2020wiy,Domingo:2021kud,Domingo:2021jdg}. More
specifically, we considered the fermionic decays of heavy,
doublet-dominated Higgs bosons in \citere{Domingo:2019vit}, stressing
the impact of (non-divergent) infrared~(IR) effects in the electroweak
corrections and pointing at the necessity of including (off-resonance)
three-body decays at tree level for a consistent analysis of the
branching ratios and widths of two-body decays at the full
  one-loop order. In \citeres{Domingo:2020wiy,Domingo:2021kud}, we
further studied the conditions for a predictive inclusion of radiative
corrections in Higgs mass and decay calculations, avoiding large
spurious effects associated with an artificial dependence on
gauge-fixing parameters and field renormalization, as commonly found
in earlier estimates. In \citere{Domingo:2021jdg}, we examined the
Higgs phenomenology in the presence of very light singlet states,
showing in particular that a tachyonic tree-level spectrum does not
necessarily yield an unphysical point in parameter space, but may
simply correspond to a problematic choice of renormalization
scheme. With the current paper focusing on 
bosonic decays of heavy, doublet-like
  Higgs states, we conclude this overview of radiative corrections
applying to Higgs decays into SM~final states in the~(N)MSSM. Similar
projects were presented in
\citeres{Graf:2012hh,Muhlleitner:2014vsa,Goodsell:2014pla,Goodsell:2015ira,Goodsell:2016udb,Goodsell:2019zfs,Dao:2021khm,Goodsell:2017pdq,Belanger:2017rgu,Baglio:2019nlc,Dao:2020dfb,Braathen:2021fyq}.

Our implementation of the bosonic Higgs decays narrowly follows the
steps outlined in
\citeres{Domingo:2019vit,Domingo:2020wiy,Domingo:2021kud,Domingo:2021jdg}
and we simply discuss the consequences of the 1L~corrections to these
channels for the Higgs phenomenology in two scenarios illustrative of
the~MSSM and the~NMSSM. In particular, we attempt to quantify
higher-order uncertainties, which may remain considerable for some
channels at~1L due to a suppressed tree-level width. We first
consider bosonic Higgs decays in the~MSSM in \sect{sec:MSSM},
insisting on the implications of the approximate
$SU(2)_\mathrm{L}$~symmetry for both two-body and three-body
decays. We then turn to the $Z_3$-conserving~NMSSM in
\sect{sec:NMSSM}, showing how channels with singlet scalars in the
final state may come to dominate the total width. We briefly summarize
the achievements of this paper in \sect{sec:conclusions}.

\section{The bosonic Higgs decays in the MSSM \label{sec:MSSM}}

We first focus on the status of bosonic decays of heavy Higgs states
in the~MSSM. We denote the \CP-even Higgs states as
\mbox{$h=-s_{\alpha}\,h_d^0+c_{\alpha}\,h_u^0$} and
\mbox{$H=c_{\alpha}\,h_d^0+s_{\alpha}\,h_u^0$}, the \CP-odd one as
\mbox{$A=s_{\beta}\,a_d^0+c_{\beta}\,a_u^0$}, and the charged ones as
$H^{\pm}=s_{\beta}\,H_d^{\pm}+c_{\beta}\,H_u^{\pm}$, with transparent
notations for the components of the doublets $H_{d}$, $H_u$,
respectively taking a vacuum expectation value (v.e.v.)
$v\,c_{\beta}$, $v\,s_{\beta}$, with
$v\equiv(2\,\sqrt{2}\,G_F)^{-1/2}$ and~$G_F$ denoting the
Fermi constant. The lightest state~$h$ is identified with the
SM-like Higgs boson observed at the~LHC. We will employ the
notation~$M_{\text{EW}}$ to represent any mass of electroweak
proportions, \IE~$M_{W,Z,h,t}\sim M_{\text{EW}}$. The masses
of all heavy doublet-Higgs states scale like~$M_{H,A}\sim
M_{H^{\pm}}$, which we will regard as falling above~$0.5$\,TeV.

\subsection{Two-body decays}

In the MSSM, the squared-mass splitting among members of the heavy
Higgs doublet ($H$, $A$, $H^{\pm}$) is of order~$M_{\text{EW}}^2\sim
M^2_{Z,W}$, implying a quasi-degeneracy at large~$M_{H^{\pm}}$. This
is a consequence of the restored $SU(2)_\mathrm{L}$~symmetry at
energies sufficiently far above the electroweak-breaking
scale. Correspondingly, the accessible decay channels of these doublet
scalars into pairs of SM~bosons are restricted to
\begin{itemize}
\item
decays into pairs of gauge bosons: $H,\,A\to
gg,\,\gamma\gamma,\,Z\gamma,\,ZZ,\,W^+W^-$; $H^{\pm}\to
ZW^{\pm},\,\gamma W^{\pm}$;
\item
decays into a pair of SM-like Higgs bosons: $H\to hh$;
\item
decays into a SM-like Higgs and an electroweak gauge boson: $A\to Zh$;
$H^{\pm}\to W^{\pm}h$;
\end{itemize} 
where we have neglected a possible \CP-violating mixing, which would
combine the decay modes of~$H$ and~$A$. All these channels have in
common to violate the $SU(2)_\mathrm{L}$~symmetry, resulting in a
suppression of the associated widths by the
ratio~$M_{\text{EW}}/M_{H^{\pm}}$ at large~$M_{H^{\pm}}$.
Correspondingly, all tree-level amplitudes involving a massless gauge
boson vanish. In the case of~$H\to hh$, the trilinear Higgs coupling
is directly proportional to the electroweak v.e.v.~$v$. For other
channels, the tree-level amplitudes are mediated by the
$SU(2)_\mathrm{L}$-breaking mixing angle in the \CP-even
sector~$\alpha-\beta+\pi/2=\mathcal{O}\big(M_{\text{EW}}^2/M^2_{H^{\pm}}\big)$.
Given that the suppression of these amplitudes in the large-mass limit
is protected by a symmetry, it is also necessarily observed by loop
corrections. For this reason, heavy-Higgs decays in the~MSSM are
dominated by the fermionic channels, which we studied in depth in
\citere{Domingo:2019vit}, while $SU(2)_\mathrm{L}$-conserving bosonic
decay modes, such as~$H\to ZA$, are kinematically inaccessible and
contribute to final states of higher multiplicity, \EG~$H\to
Zb\bar{b}$, after interfering with other diagrams.

\begin{figure}[p!]
\centering
\includegraphics[width=\linewidth]{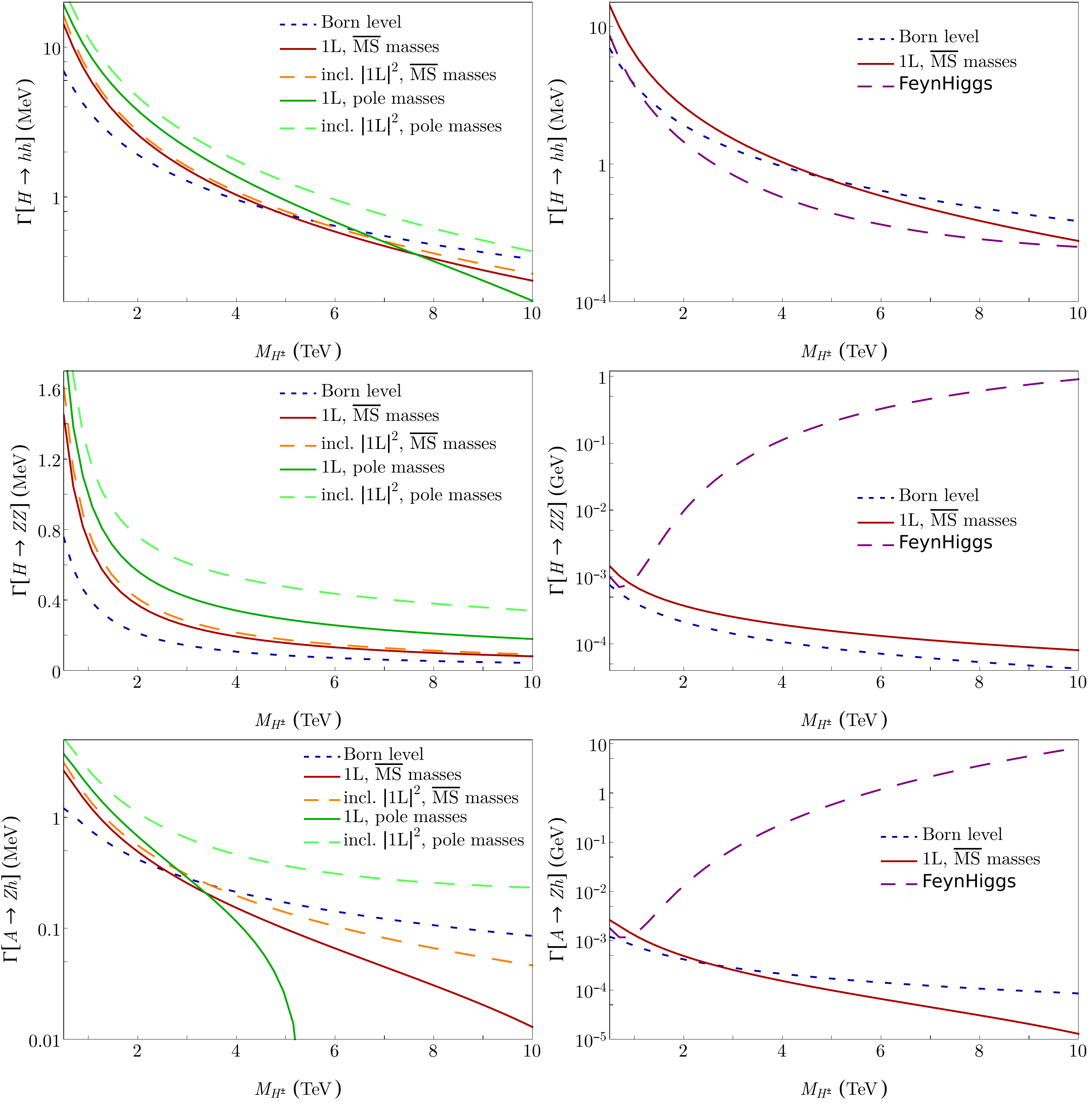}
\caption{Bosonic two-body decay widths as a function of the heavy
  Higgs scale in the MSSM at $t_{\beta}=10$: $\Gamma[H\to hh]$,
  $\Gamma[H\to ZZ]$ and $\Gamma[A\to Zh]$ are displayed in the first,
  second and third row, respectively.
\newline {\em Left}: the born-level amplitude is shown in short-dashed
blue, the strict 1L amplitude in solid lines, while the long-dashed
curves include a $\text{1L}^2$ term. In red, quark masses are set to
the running $\overline{\text{MS}}$ value at the Higgs-mass scale; in
green, pole masses are employed.
\newline {\em Right}: The prediction of \texttt{FeynHiggs} in dashed
purple is compared to a selection of curves of the plot on the
left-hand side. \label{fig:H2bos}}
\end{figure}

We compute radiative corrections to the decay amplitudes at full
1L~order, taking care---by the means of a strict perturbative
expansion---not to introduce $SU(2)_\mathrm{L}$-breaking artifacts in
the calculation\,\cite{Domingo:2020wiy}. We will not narrowly
investigate the pure radiative decays here: our implementation has
been described in \citere{Domingo:2021jdg}---see also
\citere{Domingo:2018uim}---and includes known higher-order
QCD~corrections to the diphoton\,\cite{Spira:1995rr} and
digluon\,\cite{Baikov:2006ch,Muhlleitner:2006wx} widths---see also the
summary in \citere{Spira:2016ztx}. Instead, we focus on final states
involving weak gauge or lighter Higgs bosons. A few channels are
displayed for illustration in \fig{fig:H2bos}. The SUSY~spectrum is
decoupled with masses at~$10$\,TeV while~$t_{\beta}=10$ (see
\appx{ap:input} for a summary of the input parameters). At the tree
level, the considered decay widths read:
\begin{subequations}\label{eq:bornH2b}
\begin{align}
  \Gamma[H\to hh] &= \frac{\lvert g_{Hhh}\rvert^2}{32\,\pi\,M_H}\,
  \sqrt{1-\tfrac{4\,M_h^2}{M_H^2}}\,, &
  g_{Hhh} &\stackrel[\alpha\to\beta-\tfrac{\pi}{2}]{}{\approx}
  \frac{3\,M_Z^2}{2\,\sqrt{2}\,v}\,s_{4\beta}\sim M_{\text{EW}}\,,\taghere[-3ex]\\[1ex]
  \Gamma[H\to ZZ] &= \frac{M_H^3\,\lvert g_{HZZ}\rvert^2}{128\,\pi\,M_Z^4}\,
  \Big[1-\tfrac{4\,M_Z^2}{M_H^2}+\tfrac{12\,M_Z^4}{M_H^4}\Big]\,
  \sqrt{1-\tfrac{4\,M_Z^2}{M_H^2}}\,, &
  g_{HZZ} &= \frac{\sqrt{2}\,M_Z^2}{v}\,c_{\alpha-\beta}\sim
  \frac{M^3_{\text{EW}}}{M_{H^{\pm}}^2}\,,\\
  \Gamma[A\to Zh] &= \frac{M_A^3\,\lvert g_{AZh}\rvert^2}{64\,\pi\,M_Z^2}
  \left[1-\tfrac{2\left(M_h^2+M_Z^2\right)}{M_A^2}+
    \tfrac{\left(M_h^2-M_Z^2\right)^2}{M_A^4}\right]^{3/2},&
  g_{AZh} &= \sqrt{g_1^2+g_2^2}\,c_{\alpha-\beta} \sim
  \frac{M^2_{\text{EW}}}{M_{H^{\pm}}^2}\,.
\end{align}
\end{subequations}
As expected, all these decay widths scale
like~$M^2_{\text{EW}}\big/M_{H^{\pm}}$ in the limit of large doublet
masses. The born-level amplitude for the $H\to hh$~decay does not
vanish in the limit~\mbox{$\alpha\to\beta-\tfrac{\pi}{2}$} and is
merely $t_{\beta}$-suppressed. On the contrary,~$H\to ZZ$ and~$A\to
Zh$ become purely radiative in this limit. For these latter channels,
the alignment of the \CP-even sector in the
limit~\mbox{$M_{H^{\pm}}\gg M_{\text{EW}}$} is crucial to recover the
correct behavior.

In the left column of \fig{fig:H2bos}, the tree-level prediction
(short-dashed blue lines) for the channels of \refeqs{eq:bornH2b} is
shown against~$M_{H^{\pm}}$ and compared to our estimates at~1L (red
or green solid curves). All curves fall at large~$M_{H^{\pm}}$, as
expected for $SU(2)_\mathrm{L}$-violating channels. We observe that
the magnitude of 1L~contributions is comparable to that of the
born-level amplitudes, indicating that the perturbative series
converges slowly. In these conditions, the inclusion of a
$\lvert\text{1L}\rvert^2$~term (long-dashed curves) seems numerically
justified. Nevertheless, such an operation is in general
dangerous\,\cite{Domingo:2020wiy}, because the
$\lvert\text{1L}\rvert^2$~term is a partial 2L~contribution which
generally contains loose UV-logarithms or uncontrolled
symmetry-violating contributions, thus spoiling a quantitative
interpretation. Yet, in the case of channels that are purely radiative
in the alignment limit, the 1L~amplitude is purged of its undesirable
features when explicitly applying the
limit~\mbox{$\alpha\to\beta-\tfrac{\pi}{2}$}: the corresponding piece
typically dominates at the numerical level, thus endowing the
$\lvert\text{1L}\rvert^2$~term with some degree of legitimacy. In any
case, comparing the strict 1L~widths with their counterparts including
a $\lvert\text{1L}\rvert^2$~term provides us with a first assessment
of the higher-order uncertainty, which reaches as much as~$100\%$ for
the considered channels.

Another measurement of the large uncertainties at stake at 1L~order
can be derived from the variation of input in the quark sector: quark
masses indeed enter the calculation only in 1L~diagrams---both in
triangles and self-energy corrections on the external legs---so that a
choice of scheme input appears as a higher-order concern. Thus, we
compare an on-shell input (green colors) and an
$\overline{\text{MS}}$~input with QCD~running at the scale of the
heavy Higgs masses (red/orange shades). Once again, the variation
reaches a $100\%$~magnitude, due to the substantial weakening of
Yukawa interactions at high energy. Specializing on the example
of~$H\to hh$, the contribution of the top quark to the 1L~amplitude is
of
order~$g_{Hhh}^{(\text{1L}),t}\approx3\big/\big(16\,\sqrt{2}\,\pi\big)\,m_t^4\big/(v^3\,t_{\beta})\ln^2\big(M_{H^{\pm}}^2\big/m_t^2\big)$,
implying a quartic dependence on the quark mass input, hence a
dramatic dependence on QCD~corrections of higher order. In practice,
the choice of $\overline{\text{MS}}$~masses at the high scale is
expected to properly resum the associated logarithms of UV~type, and
we correspondingly choose it as our prediction per
default. Nevertheless, our discussion should emphasize the
impossibility to claim accurate results in the prediction of the
bosonic Higgs decays in a 1L~analysis: all that is achieved is setting
the order of magnitude of the suppressed two-body widths---which is
anyway constrained by the symmetries and dimensional analysis.

Despite these bleak conclusions as to the possibility of
quantitatively exploiting the bosonic two-body decays of a heavy
doublet-Higgs state in the~MSSM---they are
suppressed and imprecise at 1L---it is crucial to properly estimate
their magnitude. On the right-hand side of \fig{fig:H2bos}, still for
the channels~$H\to hh$, $H\to ZZ$ and~$A\to Zh$, we display the tree-level (short dashed blue line) and
1L~(solid red line) estimates---identical to the corresponding curves
on the left plots---and compare them to the predictions of
\texttt{FeynHiggs-2.18.1}\,\cite{Heinemeyer:1998yj,Heinemeyer:1998np,Degrassi:2002fi,Frank:2006yh,Hahn:2013ria,Bahl:2016brp,Bahl:2017aev,Bahl:2018qog}
(long dashed purple curves). \texttt{FeynHiggs} employs a partial
1L~description of the considered widths, where the tree-level
amplitude is rotated with an effective mixing matrix diagonalizing the
Higgs-mass system at 1L~(and leading
2L)~order\,\cite{Williams:2011bu}. In the case of~$H\to hh$, this
approach does not particularly improve the prediction with respect to
a tree-level description (as it is not closer to the full 1L~result)
but, in view of the sizable uncertainties at stake, it does not
notably worsen it either, with, in particular, the correct decoupling
behavior at high mass.

However, if we turn to~$H\to ZZ$, we observe a problematic growth of
the width predicted by \texttt{FeynHiggs} for~$M_{H^{\pm}}\gg
M_{\text{EW}}$. This response is the direct consequence of the
inclusion of a partial 1L~order containing
$SU(2)_\mathrm{L}$-violating pieces that are not controlled by the
electroweak~v.e.v.: non-decoupling mixing effects actually need to be
carefully balanced with the vertex piece---which obviously fails if
the latter is not included. We already described this issue in
\citere{Domingo:2020wiy} (see \EG~Fig.\,10 in this reference). At this
point, one may wonder why the same cause does not result in
similarly devastating consequences for~$H\to
hh$: the reason simply rests with the fact that, in this latter case,
the $SU(2)_\mathrm{L}$-breaking parameter is already contained within
any of the (tree-level) triple-Higgs couplings, \IE~$g_{hhh}$
and~$g_{Hhh}$ both scale like~$M_{\text{EW}}$; on the contrary,
for~$H\to ZZ$, the $hZZ$~coupling is unsuppressed, so that a
non-decoupling behavior emerges from a call to this quantity via a
(non-decoupling) effective mixing. A similar issue appears in~$A\to
Zh$, with the difference that the non-decoupling mixing is introduced
at the level of the SM-like state, through a call to the
$SU(2)_\mathrm{L}$-conserving $AZH$~coupling.

Finally, we warn the reader against computing off-shell decays~$H\to
Z^*A^*$, $A\to Z^*H^*$, etc., as an integration over off-shell momenta
of the final state bosons, after the fashion of Eq.\,(117) of
\citere{Spira:2016ztx}. Such expressions are meant to estimate the
contribution to four-fermion final states, but the latter involve
several destructively interfering diagrams in the case of heavy
doublet-Higgs states---with \EG~$Z$-boson emission from a fermion
instead of a Higgs line. In our approach, corresponding effects are
described in terms of three-body decay widths, \EG~$A\to
f\bar{f}\,Z,\, f\bar{f}\,h$, including all tree-level contributions
and resumming Sudakov double~logarithms (which appear in phase-space
integrals)\,\cite{Domingo:2019vit}: this results in a less dramatic
behavior at high mass, consistent with the
$SU(2)_{\mathrm{L}}$~symmetry (contrarily to the off-shell
approach). In the case of decays above threshold, such as~$A\to Zh$,
use of the integration formula is possible, since the on-shell
contribution dominates the integral; on the other hand, it also brings
little, far from threshold, as compared to a simple on-shell
calculation.

\subsection{Three-body decays \label{sec:MSSM3b}}

Order counting sets a three-body width at tree level on the same
footing as 1L~contributions to a two-body decay. When enough phase
space is available, it is thus impossible to perform a complete
1L~analysis without studying three-body widths---even after discarding
possible contributions with an on-shell internal line, which we count
as two~body. In the case of fermionic decays, we have shown in
\citere{Domingo:2019vit} how IR~effects (Sudakov~logarithms)
compensate between three-body and two-body channels, resulting in
sizable effects when assessing the full width or branching
ratios. Here we focus on final states that strictly involve bosonic
particles. For the~MSSM, the relevant channels are
\begin{itemize}
\item
three gauge bosons: $H,\,A\to ZW^+W^-$; $A\to ZZZ$; $H^{\pm}\to
W^{\pm}W^+W^-,\,W^{\pm}ZZ$;
\item
two gauge bosons and one Higgs boson: $H,\,A\to hW^+W^-$; $H\to hZZ$;
$H^{\pm}\to hW^{\pm}Z$;
\item
one gauge boson and two Higgs bosons: $A\to hhZ$; $H^{\pm}\to hh W^{\pm}$;
\item
three Higgs bosons: $H\to hhh$;
\end{itemize}
where we have neglected \CP-violating mixing in the neutral sector
(indeed absent at tree~level) as well as photon radiation---which we
include in the two-body widths so that these are IR-safe with respect
to~QED.

Contrarily to the two-body decays, many of these channels are
$SU(2)_{\mathrm{L}}$-conserving---it is indeed possible to build a
doublet out of \EG~three doublets, or one doublet and two
triplets. Therefore, no suppression associated to the gauge group
arises at the amplitude level, and these channels only entail a
three-body phase-space suppression, as well as a
$t_{\beta}$-suppression: the corresponding widths then scale linearly
with~$M_{H^{\pm}}$. For example, in the limit~$M_{H^{\pm}}\gg
M_{\text{EW}}$, the $H\to hhh$~width is dominated by the contribution
of the quartic Higgs coupling:
\begin{align}
  \Gamma[H\to hhh] &\stackrel[M_{H^{\pm}}\gg M_{\text{EW}}]{}{\approx}
  \frac{\lvert g_{Hhhh}\rvert^2\,M_H}{3072\,\pi^3}\,, &
  g_{Hhhh} &\stackrel[\alpha\to\beta-\tfrac{\pi}{2}]{}{\approx}
  \frac{3}{8}\left(g_1^2+g_2^2\right)s_{4\beta}\,.
\end{align}
Due to this behavior, the three-body widths involving bosonic final
states overtake their two-body counterparts as soon as~$M_{H^{\pm}}$
reaches a few~TeV.

\begin{figure}[b!]
\centering
\includegraphics[width=\linewidth]{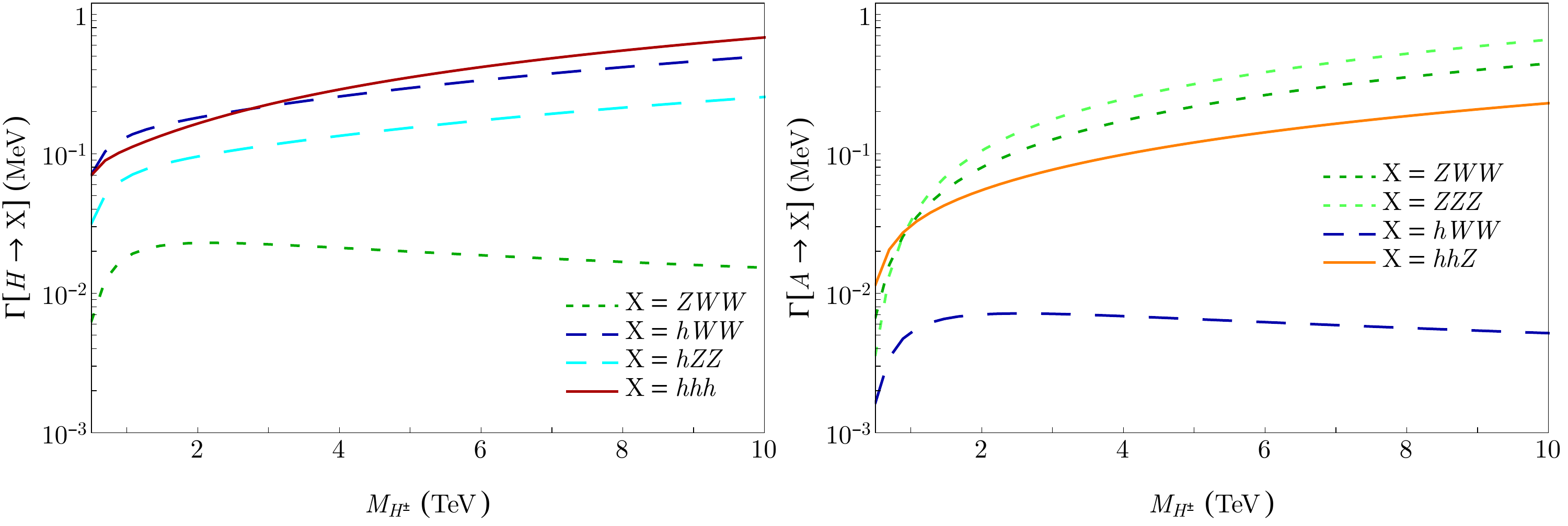}
\caption{Bosonic three-body decay widths as a function of the heavy
  Higgs scale in the~MSSM at~$t_{\beta}=10$: the non-trivial channels
  are plotted in the cases of a neutral \CP-even (left) and \CP-odd
  Higgs (right). \label{fig:H3bos}}
\end{figure}

The decays of the neutral heavy-doublet states are shown for
illustration in \fig{fig:H3bos}. For each state, \CP-even or odd,
three channels are $SU(2)_{\mathrm{L}}$-conserving and the associated
widths grow linearly with the Higgs mass, while one-channel is
$SU(2)_{\mathrm{L}}$-violating, hence suppressed---\CP-violating
channels vanish. One can easily identify the status of each channel
with respect to the $SU(2)_{\mathrm{L}}$~symmetry by checking the
existence of a corresponding quartic coupling in the tree-level
lagrangian; in such a game, gauge bosons should be identified with
their associated Goldstone bosons. Furthermore, the behavior at
large~$M_{H^{\pm}}$ is mostly determined by
these quartic couplings and one deduces the following relations in
this limit:
\begin{alignat}{2}
  \Gamma[H\to hhh] &\approx \frac{3}{2}\,\Gamma[H\to hWW] &&\approx
  3\,\Gamma[H\to hZZ]\notag\\
  \approx\Gamma[A\to ZZZ] &\approx\frac{3}{2}\,\Gamma[A\to ZWW] &&\approx
  3\,\Gamma[A\to hhZ]\,.
\end{alignat}

\begin{figure}[p!]
\centering
\includegraphics[width=\linewidth]{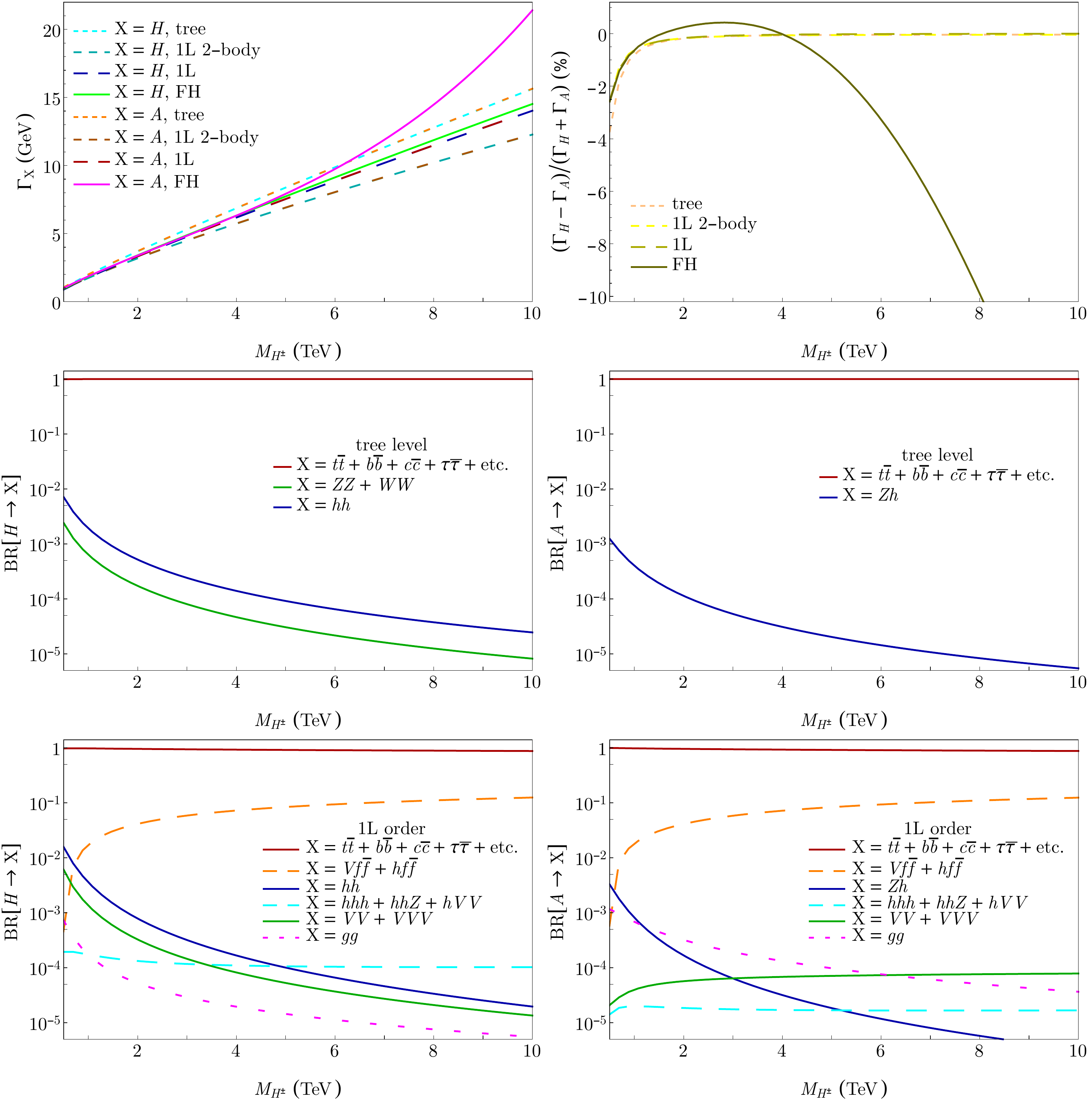}
\caption{Total widths, magnitude of $SU(2)_{\mathrm{L}}$-breaking, and
  branching ratios of the neutral heavy Higgs bosons at~$t_{\beta}=10$
  for the~MSSM with a decoupled SUSY~sector.
  \newline{\em Top Left}: Total widths at (QCD+QED-corrected) tree
  level (short-dashed curves), full~1L (long-dashed curves) and
  pseudo-widths (sum of 1L-corrected two-body widths; in intermediate
  dashes). The predictions of \texttt{FeynHiggs} are shown as solid
  lines.
  \newline{\em Top Right}: Magnitude that
  $SU(2)_{\mathrm{L}}$-breaking reaches at the level of the full
  (pseudo)widths in the various approaches.
  \newline{\em Middle}: Branching ratios at tree level: fermionic
  (red), Higgs (blue) and gauge-boson (green) two-body channels.
  \newline{\em Bottom}: Branching ratios at~1L: three-body fermionic
  (dashed orange) and bosonic (involving one SM-like Higgs; in dashed
  cyan) modes, as well as the two-body gluonic rate (dashed magenta)
  are added to the picture.\label{fig:BRs}}
\end{figure}

The theoretical uncertainty in all these channels can be expected to
be sizable (of order~$100\%$), since the widths are evaluated at
leading order---consistently with the 1L~order for two-body
decays---and the Higgs potential is known to receive sizable radiative
corrections from \EG~Yukawa interactions. Therefore, as for the
two-body decays, an evaluation at the considered order only fixes the
general magnitude of the three-body widths, without allowing for
precision tests.

The decays of the charged Higgs bosons are constrained by the
$SU(2)_{\mathrm{L}}$~symmetry to follow a pattern similar to the one
of the neutral states, hence these channels exhibit limited
novelty. For completeness, we collect a few plots in
\appx{ap:Hpbosdec}.

\subsection{Branching ratios}

Having estimated all Higgs decays into SM~two-body final states at~1L
and three-body final states at tree~level, it is possible to
consistently evaluate the full widths and branching ratios at this
order in scenarios where exotic particles~(SUSY) are kinematically
inaccessible.

The total widths for the heavy neutral states at~$t_{\beta}=10$ are
shown in the first row of \fig{fig:BRs}. As expected from the
$SU(2)_{\mathrm{L}}$~symmetry, the predicted widths are comparable for
both states, either at the tree~level (short-dashed lines) or at the
loop level (long-dashed lines). The results of \texttt{FeynHiggs}
(solid curves) do not satisfy this property at large~$M_{H^{\pm}}$,
due to the issues mentioned earlier in the assessment of bosonic decay
widths, mainly. In addition, the apparent agreement of the
expectations of \texttt{FeynHiggs} for the \CP-even state (in green)
with our full 1L~result is in fact coincidental. Given that
\texttt{FeynHiggs} does not consider three-body decays, its
predictions should be compared to the pseudo-widths, where only
1L~two-body widths are summed (not IR-safe with respect to weak
corrections; plotted in dashes of intermediate size): the mismatch is
of order~$15\%$ and can be explained, in part by the problematic $H\to
ZZ,WW$~widths, but also by the contamination of the fermionic widths,
on the side of \texttt{FeynHiggs}, by symmetry-violating terms of
higher order---we refer the reader to the more detailed discussion of
\citere{Domingo:2020wiy}.

\enlargethispage{.1ex}
The branching ratios at full 1L~order are shown in the last row of
\fig{fig:BRs} and can be compared to the (QCD-corrected) tree-level
version (in the middle). The main difference is related to the
emergence of the three-body fermionic decays
(reaching~$\mathcal{O}(10\%)$ at large~$M_{H^{\pm}}$; dashed orange
lines), which complement the dominant two-body fermionic channels (in
solid red). As explained in \citere{Domingo:2019vit}, these modes
essentially involve the radiation of electroweak gauge bosons from the
initial Higgs or fermionic final states and are dominated by
double~logarithms of Sudakov~type. Bosonic branching ratios (blue and
green curves) remain below percent level---and even permil,
for~$M_{H^{\pm}}\gsim2$\,TeV. The three-body bosonic decays, in
particular those modes involving at least one SM-like Higgs boson in
the final state (dashed cyan lines), overtake the bosonic two-body
decays for~$M_{H^{\pm}}\gsim4$\,TeV. The linear growth of the
associated widths with the Higgs mass ensures that such channels are
not further suppressed at large~$M_{H^{\pm}}$, contrarily to the
$SU(2)_{\mathrm{L}}$-violating two-body modes.

The picture described above remains comparatively stable under
variation of~$t_{\beta}$, with effects associating to the top Yukawa
coupling at low~$t_{\beta}$ replacing those of the bottom
Yukawa at high~$t_{\beta}$. Therefore, as a tribute
to the $SU(2)_{\mathrm{L}}$-suppression, bosonic decays of the heavy
Higgs states always represent sub-percent-level channels for masses
above~TeV in the~MSSM. Of course, if SUSY final~states become
accessible, some of them are $SU(2)_{\mathrm{L}}$-conserving and may
compete with the SM-fermionic channels.

\section{The bosonic Higgs decays in the NMSSM \label{sec:NMSSM}}

The presence of singlet degrees of freedom in the~NMSSM may noticeably
affect the phenomenology of heavy doublet-Higgs states, in particular
if these new scalars are kinematically accessible. Indeed, the heavy
Higgs bosons may then decay into bosonic final states without breaking
the $SU(2)_{\mathrm{L}}$~symmetry, hence opening channels that are
competitive with respect to the fermionic ones. Below, we denote the
singlet-dominated \CP-even and odd scalars as~$h_S$ and~$a_S$
respectively.

\subsection{Two-body decays}

When~$h_S$ and/or~$a_S$ take a mass below~$M_{H^{\pm}}$, the following
two-body decay channels may add up to the~MSSM modes:
\begin{itemize}
\item
decays into two Higgs bosons: $H\to h_Sh_S,\,a_Sa_S,\,h_Sh$; $A\to
a_Sh_S,\,a_Sh$;
\item
decays into a Higgs and a gauge boson: $H\to Za_S$; $A\to Zh_S$;
$H^{\pm}\to W^{\pm}h_S,\,W^{\pm}a_S$.
\end{itemize}
Again, we have assumed \CP-conservation for a cleaner classification,
but the inclusion of \CP-violating mixings does not fundamentally
change the situation. Among these new channels, several are again
$SU(2)_{\mathrm{L}}$-violating, in substance all those involving only
singlet scalars in the final state. However, in contrast to the~MSSM,
some other decay modes---involving one singlet and one doublet or weak
gauge boson (\IE~its Goldstone component)---are
$SU(2)_{\mathrm{L}}$-conserving and need not be small in the
limit~$M_{H^{\pm}}\gg M_{\text{EW}}$ provided the `portal' between
singlet and~MSSM sectors, \IE~the coupling~$\lambda$, is sufficiently
intense.

\begin{figure}[tbh!]
  \centering
  \includegraphics[width=\linewidth]{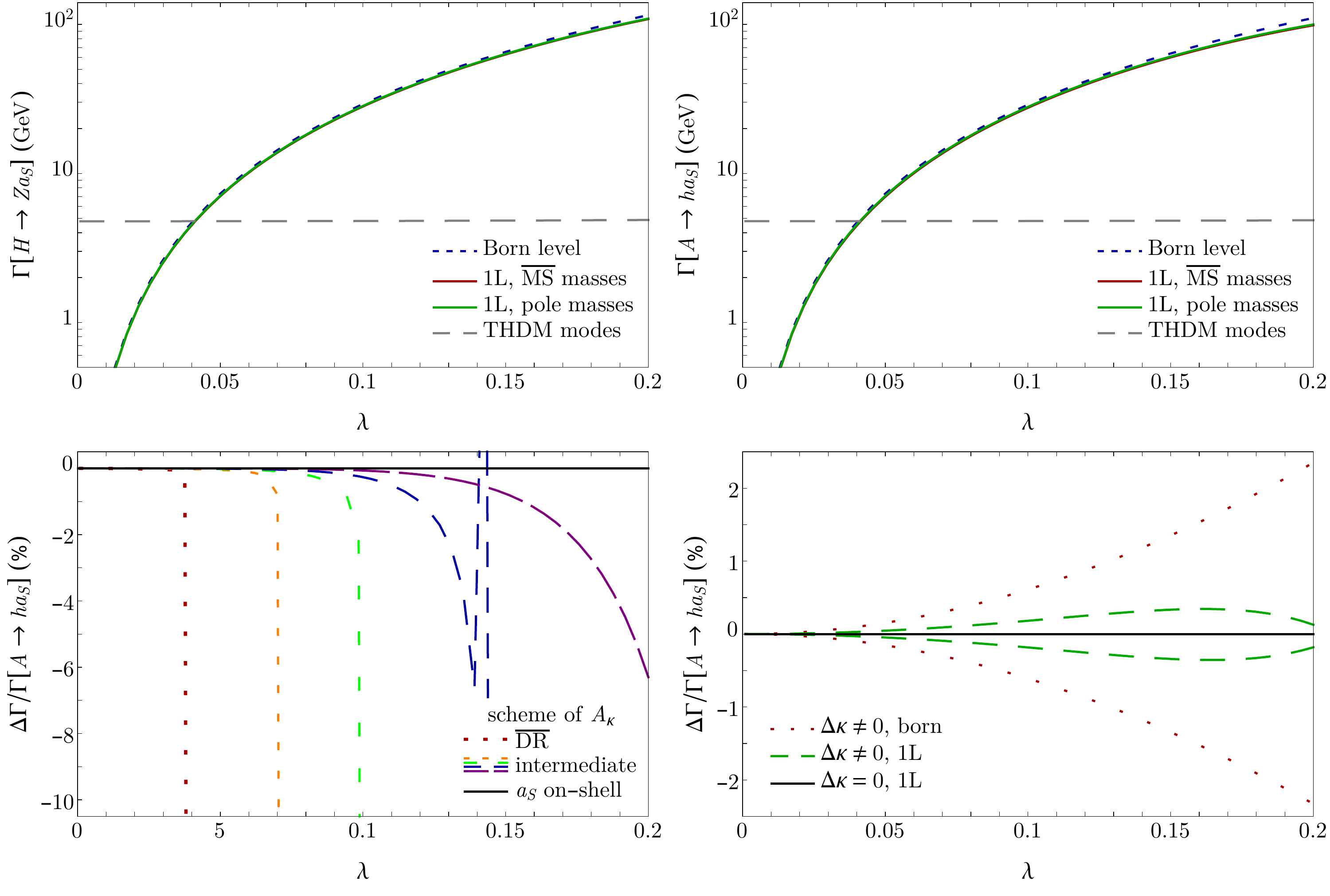}
  \caption{$SU(2)_{\mathrm{L}}$-conserving two-body decays in an NMSSM
    scenario with $\kappa\stackrel[]{!}{=}\lambda$,
    $M_{H^{\pm}}=3$\,TeV and $M_{a_S}=300$\,GeV (and $h_S$
    kinematically inaccessible).
    \newline {\em First row}: Decay widths $\Gamma[H\to Za_S]$ (left)
    and $\Gamma[A\to ha_S]$ (right) as a function of $\lambda$ at
    tree-level (blue dashed), or 1L (solid curves) with varying
    definitions of the quark masses (green vs.~red).
    \newline {\em Second row}: Behavior of $\Gamma[A\to ha_S]$ under
    variations of the renormalization scheme for $A_{\kappa}$ (left)
    or $\kappa$ (right). \label{fig:Aha}}
\end{figure}

We consider a scenario with a comparatively light \CP-odd
singlet-dominated Higgs boson, setting its mass to~$300$\,GeV,
while~$M_{H^{\pm}}\stackrel[]{!}{=}3$\,TeV. The SUSY~sector (as well
as the \CP-even singlet) remains decoupled, taking masses in
the~$10$\,TeV range. All input parameters are summarized in
\appx{ap:input}. The singlet~$a_S$ primarily decays into fermion
pairs, so that the two-body decays of the heavy doublet states that we
study below can be understood as the resonant contribution to
four-fermion decay channels. In \fig{fig:Aha}, we
vary~$\lambda=\kappa$ between low values---where NMSSM~effects are
negligible and the singlet decouples---and~$0.2$ with an active
connection between singlet and MSSM~sectors. Large radiative
corrections to the mass of the singlet pseudoscalar develop for
sizable values of~$\lambda=\kappa$: this is the simple consequence of
the quadratic dependence of scalar masses on the UV~spectrum below the
SUSY-breaking scale. Technically, this causes complications with the
requirement of~$M_{a_S}\stackrel[]{!}{=}300$\,GeV in our original
renormalization scheme (\DR~with the renormalization scale~$M_t$), as
the pseudoscalar mass may be vastly different at the tree~level and
at~1L. In fact, a `tachyonic tree-level syndrome'---see
\citere{Domingo:2021jdg}---emerges as soon as~$\lambda\gsim 0.04$,
making evaluations at the 1L~order impossible in this scheme. For this
reason, it is convenient to directly work in the renormalization
scheme where the counterterm for~$A_{\kappa}$ is fixed by an on-shell
condition on the mass of~$a_S$\,\cite{Domingo:2021jdg}, after the
tree-level input is correspondingly translated: the bare~$A_{\kappa}$
is kept unchanged in the scheme translation.

We then focus on the two $SU(2)_{\mathrm{L}}$-conserving
channels~$H\to Za_S$ and~$A\to ha_S$. At the tree~level, one may
express the widths as
\begin{subequations}\label{eq:Hhadec}
\begin{align}
  \Gamma[H\to Za_S] &= \frac{M_H^3\,\lvert g_{HZa_S}\rvert^2}{64\,\pi\,M_Z^2}
  \left[1-\tfrac{2\left(M_Z^2+M_{a_S}^2\right)}{M_H^2}+
   \tfrac{\left(M_Z^2-M_{a_S}^2\right)^2}{M_H^4}\right]^{3/2},\label{eq:HZaSborn}\\
  \Gamma[A\to ha_S] &= \frac{\lvert g_{Aha_S}\rvert^2}{16\,\pi\,M_A}
  \left[1-\tfrac{2\left(M_h^2+M_{a_S}^2\right)}{M_A^2}+
    \tfrac{\left(M_h^2-M_{a_S}^2\right)^2}{M_A^4}\right]^{1/2}.
\end{align}
\end{subequations}
In the $SU(2)_{\mathrm{L}}$~limit with~$M_{\text{EW}}\ll
M_{H^{\pm}}$, one can directly work out
\begin{align}
  g_{Aha_S} &\approx \frac{1}{\sqrt{2}}\,\Big[3\,\kappa\,\mu_{\text{eff}}-
    \tfrac{\lambda\,s_{2\beta}}{2\,\mu_{\text{eff}}}\,M_{H^{\pm}}^2\Big]\,.
\end{align}
Obviously,~$g_{Aha_S}$ does not vanish and may reach sizable
values for~$\kappa,\lambda=\mathcal{O}(1)$. This approximation is less
immediate for~$g_{HZa_S}$:
\begin{align}
g_{HZa_S} &= \sqrt{g_1^2+g_2^2}\,\Big[X^R_{Hd}\,X^I_{a_Sd}-X^R_{Hu}\,X^I_{a_Su}\Big]
\end{align}
with~$X^{R,I}_{h_id,u}$ denoting the components of the Higgs
field~$h_i$ with respect to the scalar and pseudoscalar doublet gauge
eigenstates; the gauge interaction couples only doublet, not singlet
states. Thus, the $H$--$Z$--$a_S$~coupling necessarily emerges from
singlet--doublet mixing and is $SU(2)_{\mathrm{L}}$-violating. Yet,
the prefactor~$M_Z^{-2}$ in \refeq{eq:HZaSborn} shows that the
relevant object to consider in the $SU(2)_{\mathrm{L}}$~limit is
not~$g_{HZa_S}$ but~$g_{HZa_S}\big/M_Z$. Then, it is convenient to use
the connection between gauge and Goldstone couplings, providing the
relation~\mbox{$\lvert g_{HZa_S}\rvert\big/M_Z=2\,\lvert
  g_{HG^0a_S}\rvert\big/(M_H^2-M_{a_S}^2)$}, and to apply the
$SU(2)_{\mathrm{L}}$~limit to the Goldstone coupling, \IE
\begin{align}
  g_{HG^0a_S} &\approx -\frac{1}{\sqrt{2}}\,
  \Big[3\,\kappa\,\mu_{\text{eff}}-\tfrac{\lambda\,s_{2\beta}}{2\,\mu_{\text{eff}}}\,
    M_{H^{\pm}}^2\Big]\,.
\end{align}
At this point, we see that the $H\to Za_S$~transition actually amounts
to a Higgs-to-Higgs decay in the $SU(2)_{\mathrm{L}}$~limit, the
$Z$~boson serving as proxy for its Goldstone counterpart. In addition,
both channels of \refeq{eq:Hhadec} are related by the
$SU(2)_{\mathrm{L}}$~symmetry, so that the widths are approximately
equal. Radiative corrections are expected to abide by such a
symmetry-protected relation.

The decay widths are displayed in the first row of \fig{fig:Aha} at
the tree~level (blue short-dashed line) and at~1L for different
definitions of the Yukawa couplings (solid green and red curves). The
width corresponding to the sum of the MSSM-like decay modes is also
shown (long-dashed gray line) and remains comparatively stable with
varying~$\kappa=\lambda$. Contrarily to the MSSM~case, Higgs-to-Higgs
decays are found to dominate over the fermionic modes,
provided~$\lambda=\kappa$ is sufficiently large. The reason
  for this behavior is the
dependence of~$g_{Aha_S,HG^0a_S}$ on the scale~$\kappa\,\mu_{\text{eff}}$, which can compete
with~$Y_f\,M_{H^{\pm}}$ in the fermionic case (with~$Y_f$ denoting the
Yukawa coupling of the fermion species~$f$). Now, focusing on the
bosonic decay widths, we observe that these are largely determined by
the tree-level amplitudes; 1L corrections
only amount to~$\mathcal{O}(10\%)$. This again
contrasts with the MSSM-like bosonic channels. A fast convergence of
the perturbative series is expected and the inclusion
of~$\lvert\text{1L}\rvert^2$~terms appears unjustified (and
unpredictive) in such a context. The choice of input for the fermion
couplings also appears largely unimportant, with an impact
of~$\mathcal{O}(1\%)$.

In the second row of \fig{fig:Aha}, we test the scheme dependence
in~$\Gamma[A\to ha_S]$. In the plot on the left, we study the impact
of the treatment of~$A_{\kappa}$, considering the variations between a
calculation in the scheme with on-shell~$M_{a_S}$ (black solid line),
the original \DR~scheme (dotted red) and several intermediate
choices. The `collapse' of the predictions for non-OS~choices is
related to the tree-level mass~$m_{a_S}$ becoming tachyonic; this
issue should not distract our attention too much: perturbative
calculations are expected to fail for an ill-defined spectrum, and the
OS~choice is \AP~more predictive. More interestingly, in the regime
where these schemes are well-defined, their predictions for the decay
width do not differ from that obtained with the OS~choice by more than
a few~percent---and this magnitude is reached only in the limit of
validity of the non-OS~schemes. This confirms the stability of the
predicted width under varying definitions of the
renormalized~$A_{\kappa}$. Yet, given the form of the tree-level
couplings---see the paragraph after \refeq{eq:Hhadec}---we expect the
main uncertainty to originate in the definition of~$\kappa$
(or~$\mu_{\text{eff}}$). Thus, we also vary the scheme associated
with~$\kappa$ in the plot on the right-hand side: keeping the
bare~$\kappa$ unchanged, we consider variations of the renormalized
parameter
by~$\Delta\kappa\approx3\big/(16\,\pi^2)\,\kappa\,(\lambda^2+\kappa^2)\ln\big(M_{\text{SUSY}}^2\big/M_t^2\big)$,
as suggested by the anomalous dimension at~1L. The widths at~1L
(dashed green curves) are affected at permil~level, at most, against
percent for the born level predictions (dotted red lines). This is
another sign of the convergence of the perturbative series and we can
conclude as to a higher-order uncertainty by at most a few~percent on
the $SU(2)_{\mathrm{L}}$-conserving channels.

\needspace{5ex}
The approximate $SU(2)_{\mathrm{L}}$~relation between the two
channels~$H\to Za_S$ and~$A\to ha_S$ is roughly satisfied in
\fig{fig:Aha}. Violating effects amount to at most~$\simord5\%$ at the tree~level and~$\simord10\%$ at~1L (in
this specific scenario). This apparently increasing magnitude of the
$SU(2)_{\mathrm{L}}$-breaking at~1L should not be over-interpreted:
indeed, at this order, three-body decays should be counted at the same
level as 1L~corrections to the two-body decays. In the next
subsection, we will find that the larger three-body decay width of~$A$
somewhat compensates the larger suppression of~$\Gamma[A\to ha_S]$ as
compared to~$\Gamma[H\to Za_S]$.

\begin{figure}[t!]
  \centering
  \includegraphics[width=0.5\linewidth]{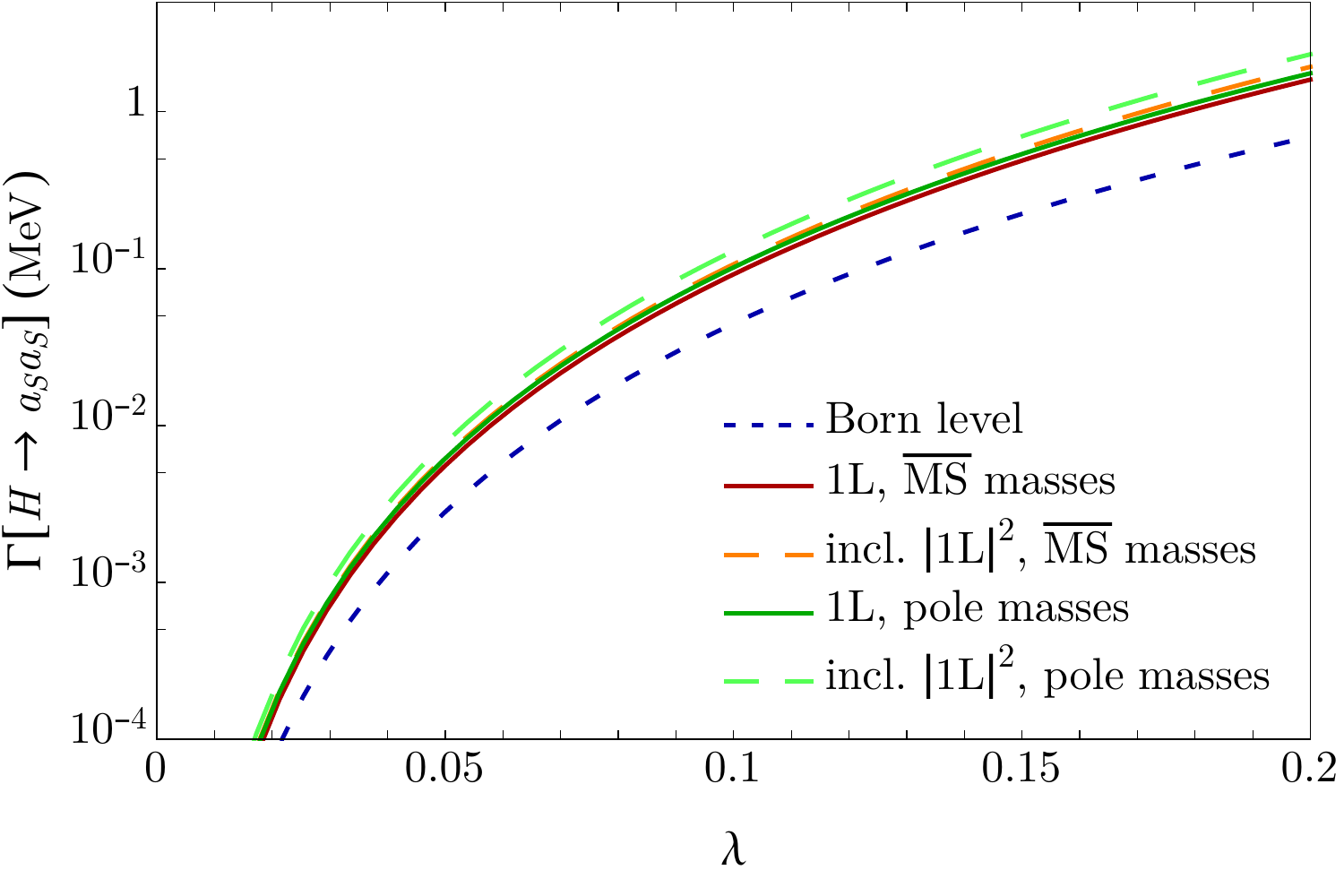}
  \caption{$SU(2)_{\mathrm{L}}$-violating two-body decay width
    $\Gamma[H\to a_Sa_S]$ in an NMSSM scenario
    with~$\kappa\stackrel[]{!}{=}\lambda$, $M_{H^{\pm}}=3$\,TeV and
    $M_{a_S}=300$\,GeV (and $h_S$ kinematically inaccessible). The
    convention for the curves is the same as in
    \sect{sec:MSSM}\label{fig:Haa}}
\end{figure}

In \fig{fig:Haa}, we consider the $SU(2)_{\mathrm{L}}$-violating
decay~$H\to a_Sa_S$ (a channel that does not exist in
  the THDM) in the same scenario as before. As
expected, this width is considerably suppressed as compared to those
of \fig{fig:Aha}, though growing with~$\lambda=\kappa$. We actually
recover features that are largely similar to the MSSM-like
Higgs-to-Higgs decays, with a slow convergence of the perturbative
series and a sizable dependence on the definition of the
SM~Yukawa~couplings, all hinting at an uncertainty of the order
of~$100\%$ at~1L. On the other hand, it is always possible to tailor
scenarios where a large mixing between~$H$ and~$h_S$ ensures a sizable
decay of their admixtures into~$a_Sa_S$---exploiting the
$SU(2)_{\mathrm{L}}$-conserving decay~$h_S\to a_Sa_S$; see
\EG~\citere{Domingo:2016unq}. As the approximate degeneracy between
singlet and doublet-dominated states is accidental, such setups are
generally fine-tuned.

Attempts at a comparison with the public tools
\texttt{NMSSMCALC}\,\cite{Baglio:2013iia} and
\texttt{NMSSMCALCEW}\,\cite{Baglio:2019nlc} did not lead to
satisfactory results. First, we stress that a preliminary translation
of the input is needed, since the renormalization scheme in these
codes is different from our choice. Then, electroweak corrections do
not seem to produce noticeable IR-like effects in the predictions of
\texttt{NMSSMCALCEW}, neither in fermionic nor
bosonic two-body modes, which contrasts with our calculation. On the
other hand, contrarily to \texttt{FeynHiggs}, we did not observe any
obvious violation of the $SU(2)_{\mathrm{L}}$~symmetry in the results
of \texttt{NMSSMCALC} and \texttt{NMSSMCALCEW}. Tree-level predictions
of bosonic two-body decays are in rough agreement with ours. A deeper
understanding of the two implementations of decay widths would thus be
needed for a meaningful comparison of the results, which goes beyond
our purpose in the current paper.

\needspace{21ex}
\subsection{Three-body decays}

Numerous bosonic three-body final states open up in addition to the
THDM~ones, when a singlet-dominated (pseudo)scalar is kinematically
accessible to the decays of heavy Higgs bosons. It is again necessary
to calculate these widths at tree~level for a meaningful 1L~analysis
of the total widths and branching ratios of Higgs bosons. There exist
fermionic three-body decay channels involving~$h_S$ and~$a_S$ as well:
these are typically dominated by the emission of a fermion pair via an
off-shell~$h$, $G^0$ or~$G^{\pm}$ in the two-body topology of the
previous subsection---see \citere{Domingo:2019vit}. We do not study
them in detail below.

\begin{figure}[b!]
  \centering
  \includegraphics[width=\linewidth]{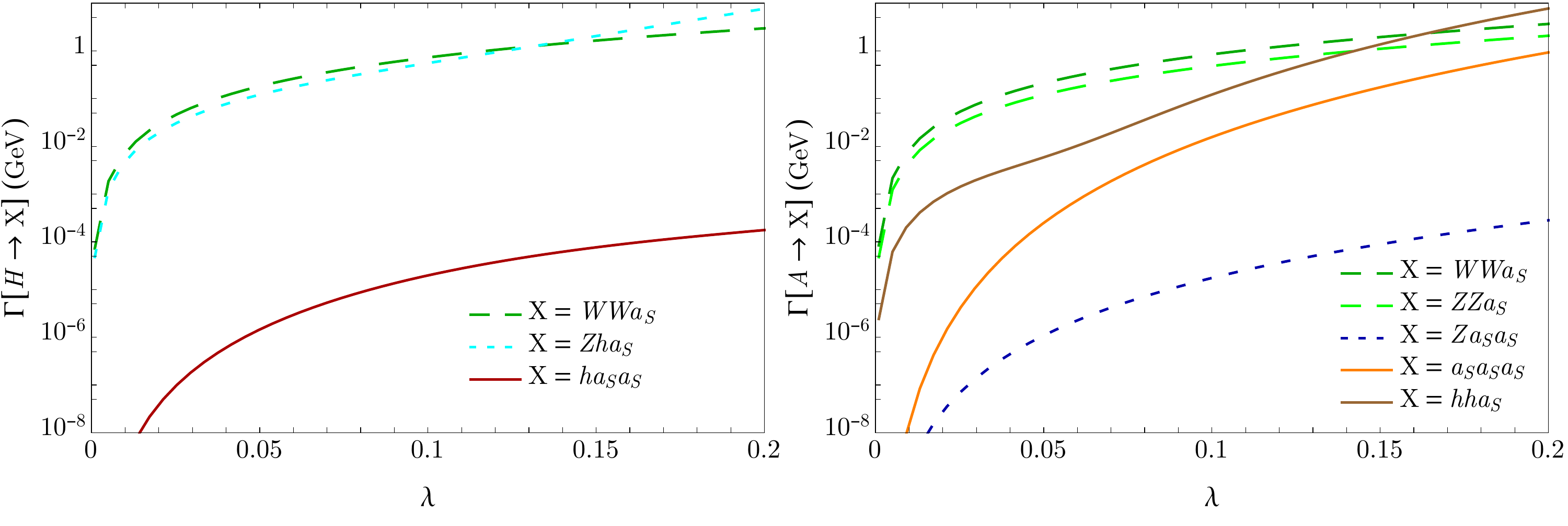}
  \caption{Bosonic three-body decay widths of the heavy doublet-like
    Higgs bosons, involving $a_S$ in the final state, in the scenario
    of \fig{fig:Aha}.\label{fig:threebNMSSM}}
\end{figure}

We consider the three-body decay widths into bosonic final states in
the scenario of \fig{fig:Aha}. As before, we exclude the contributions
from resonant internal lines, which are counted in the two-body
widths, and keep only off-shell effects at this level. We do not
discuss the modes of THDM~type here, since they continue to behave in
a fashion comparable to that observed in the~MSSM. Instead, we focus
on final states involving a pseudoscalar~$a_S$. The results are
displayed in \fig{fig:threebNMSSM}. Some channels acquire a sizable
width at large~$\lambda=\kappa$, which may actually exceed the full
width of THDM~type. In fact, as could be anticipated, the magnitude of
the bosonic three-body widths is comparable (with opposite sign) to
that of the 1L~effects in the dominant
($SU(2)_{\mathrm{L}}$-conserving) bosonic two-body widths, amounting
to~$5$--$10\%$ for this specific scenario. Given that the
1L~corrections lead to a larger suppression of the two-body width in
the pseudoscalar case, which we mentioned earlier, the three-body
widths are also more important for~$A$, as compared to~$H$.

This balance between virtual corrections and boson radiation
underlines the importance of the IR~effects in the three-body
channels. Indeed, contrarily to the THDM~three-body~modes encountered
in \sect{sec:MSSM3b}, the widths involving~$a_S$ are not characterized
by the quartic couplings, but by the `dressing' of the leading
two-body channels~$H\to Za_S$, $A\to ha_S$ and~$H^{\pm}\to W^{\pm}a_S$
with radiated light bosons. The latter evidently include electroweak
gauge bosons and explain the large contributions to \EG~$H,A\to
W^+W^-a_S,ZZa_S$. From the $SU(2)_{\mathrm{L}}$~perspective, one of
the gauge bosons then replaces the associated Goldstone boson while
the other behaves vector-like. The width~$H\to Zha_S$ could be
understood in similar terms---at least for moderate values
of~\mbox{$\lambda=\kappa$}. However, for the larger range
of~$\lambda=\kappa$, this channel shares some characteristics
with~$A\to a_Shh$, which obviously violates the
$SU(2)_{\mathrm{L}}$~symmetry, but emerges as an important
channel. The effect at stake here is the radiation of~$h$ from
the~$a_S$~line in the soft/collinear regime. Kinematical integration
in the off-shell regime close to the~$a_S$~pole indeed generates a
scaling~${\propto}\,m^{-2}_{a_S}$ of the decay width, which
compensates the suppression by the $SU(2)_{\mathrm{L}}$-breaking
trilinear Higgs coupling~$a_S$--$a_S$--$h$: in the
aftermath,~$\Gamma[A\to a_Shh]\propto M v^2/m^2_{a_S}$,
where~$M\sim\mu_{\text{eff}},M_{H^{\pm}}$ is one of the large masses
of the problem. Therefore, the importance of such
$SU(2)_{\mathrm{L}}$-violating channels emerges as a consequence from
the accidental fact that~$m^2_{a_S}\sim v^2$. Finally, we stress that,
from the IR~nature of these effects, they largely compensate between
virtual and real corrections, so that~$SU(2)_{\mathrm{L}}$ remains a
pertinent approximate symmetry for the inclusive width.

The precision on these (off-resonance) three-body decay widths
calculated at the tree~level cannot be very competitive. Given the
dominance of IR~effects, we can expect the absolute uncertainty to
match the residual one in virtual corrections to the two-body decays,
\IE~amount to percent level of the two-body widths, or~$20$--$50\%$ of
the three-body widths. Yet, the general order of magnitude should be
under control.

\subsection{Branching ratios}

\begin{figure}[p!]
  \centering
  \includegraphics[width=\linewidth]{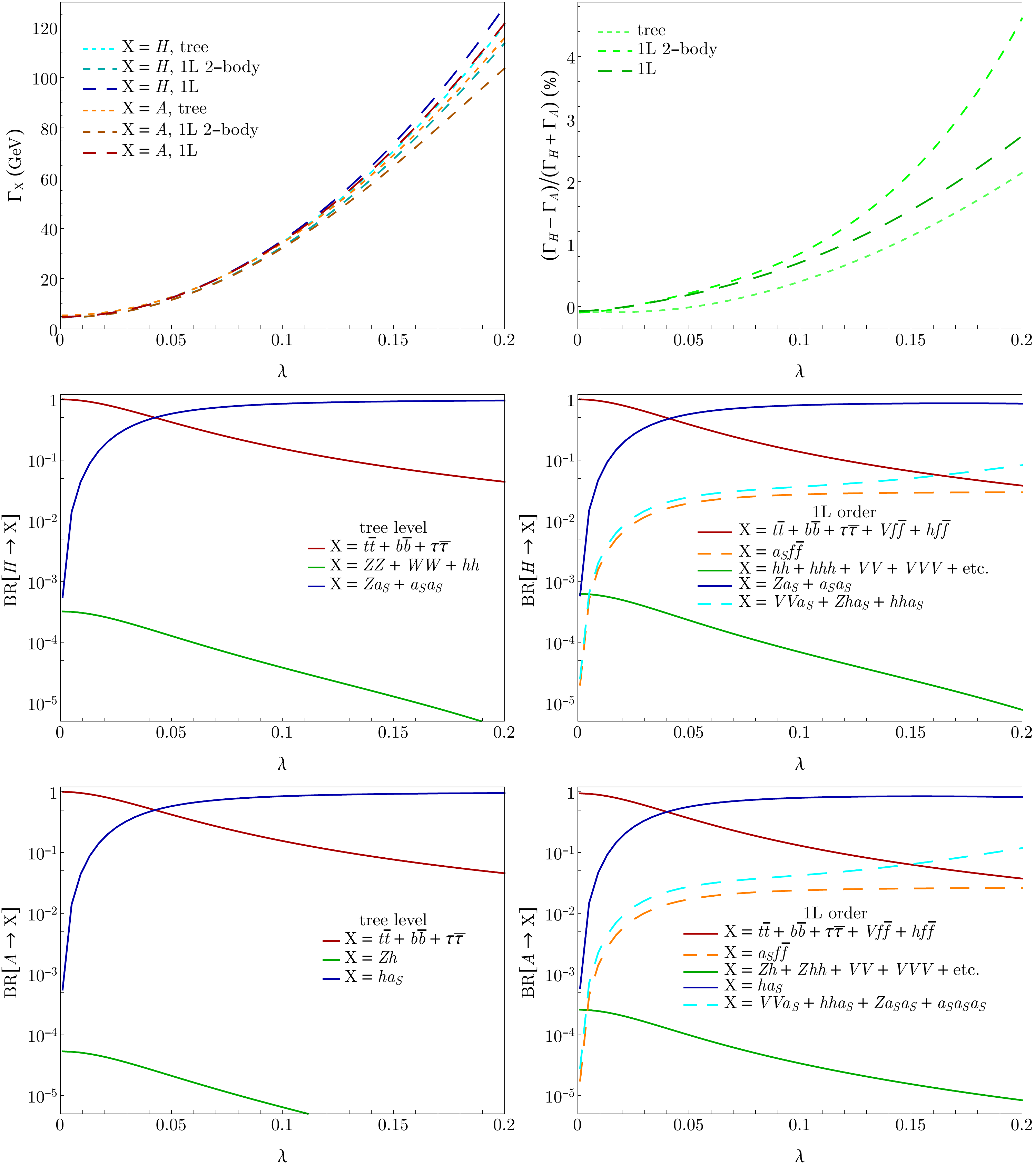}
  \caption{Total widths of the heavy doublet states (first row, left),
    magnitude that $SU(2)_{\mathrm{L}}$-breaking reaches at the level
    of the full (pseudo)widths (first row, right), and branching
    ratios (middle: \CP-even Higgs; bottom: \CP-odd) at tree-level
    (left) and 1L~order (right) in the scenario of
    \fig{fig:Aha}. Decay modes involving~$a_S$ are depicted in solid
    blue (two-body), dashed orange (fermionic three-body) and dashed
    cyan (bosonic three-body).\label{fig:BRNMSSM}}
\end{figure}

With SUSY decays kinematically inaccessible in the considered
scenario, we have completed the calculation of all the relevant widths
for a 1L~evaluation of the total widths and branching ratios of the
heavy doublet-Higgs states. The corresponding results for the scenario
of \fig{fig:Aha} are displayed in \fig{fig:BRNMSSM}.

Focusing on the total width first, we stress the relevance of the
three-body channels for a consistent evaluation: the difference
between the full widths and the pseudo-widths, considering only
two-body channels at~1L, is indeed comparable to the magnitude of the
1L~effects, implying an error of full 1L~magnitude when the three-body
modes are overlooked. In addition, we observe that the magnitude of
the $SU(2)_{\mathrm{L}}$-violating effects, measured as the difference
between the widths for the scalar and pseudoscalar states (see plot in
the top right-hand corner of \fig{fig:BRNMSSM}), remains comparable at
the tree~level and at~1L, but about doubles when considering the
pseudo-widths. This feature is the natural consequence of
$SU(2)_{\mathrm{L}}$-violating effects of IR~type developing in
virtual corrections and Higgs radiation, but compensating between the
two.

The general magnitude of the widths considerably varies with the
choice of~$\lambda=\kappa$, as NMSSM-specific effects range from
negligible to dominant. This is most visible at the level of the
branching ratios where the THDM fermionic modes (red solid curves)
decrease from a significance of about~$100\%$ at low~$\lambda$ to
percent level at~$\lambda\approx0.2$. While the absolute magnitude of
the corresponding widths (about~$5$\,GeV in this scenario) remains
constant, the competition of the bosonic Higgs decays involving~$a_S$
(blue solid line) steadily increases with~$\lambda$, resulting in a
completely different phenomenology. The general picture qualitatively
changes little between the tree~level and the 1L~order. Only the
emergence of the three-body decays diminishes the significance of the
two-body modes, substituting them with more complicated final states
for a weight of~$\simord5$--$10\%$. Both fermionic and bosonic
three-body modes involving~$a_S$ become important at large~$\lambda$
and may compete in their own right with the THDM~modes. As in the
MSSM~case, however, bosonic final states of THDM~type---\IE~involving
gauge bosons and the SM-like Higgs boson---remain subdominant in the
whole considered range of parameters.

As a concluding note, let us stress that the relative weight of the
three-body modes in the decays of heavy Higgs bosons and their mostly
IR~origin would make it necessary to closely scrutinize the
experimental cuts in order to quantitatively exploit a hypothetical
measurement of a new Higgs boson and identify the relative importance
of the various channels.

\needspace{15ex}
\section{Conclusions\label{sec:conclusions}}

In this paper, we have shown how the approximate
$SU(2)_{\mathrm{L}}$~symmetry in the (experimentally-favored) limit~$M_{H^{\pm}}\gg M_Z$ severely
constrains the decays of the heavy doublet-dominated
Higgs bosons of the~(N)MSSM. In particular, most available MSSM-like
bosonic channels involving two-body final states are purely radiative
or $SU(2)_{\mathrm{L}}$-violating, meaning that an evaluation at~1L is
only able to estimate the leading order effects, with a very large
uncertainty---due to \EG~QCD~corrections to fermion loops---persisting
in the absolute determination of the widths and branching
  ratios. The fermionic decay modes thus emerge as the leading
contributors to the widths of the heavy-doublet scalars of
the~MSSM. The bosonic three-body decays are not necessarily
$SU(2)_{\mathrm{L}}$-violating and prove competitive with the two-body
ones as soon as~$M_{H^{\pm}}$ reaches a few~TeV. Finally, comparisons
with \texttt{FeynHiggs} justify the necessity to modernize this tool
in order to study the phenomenology of heavy Higgs bosons.

The situation can drastically change in the presence of comparatively
light singlet-dominated states, provided no singlet--doublet
decoupling is enforced. Then, two-body decay modes involving
Higgs-to-Higgs transitions exist that do not break the electroweak
symmetry and may even dominate the decays of the heavy mostly-doublet
states. Since no symmetry suppresses the tree-level contributions,
such transitions are better controlled by the perturbative expansion
(in constrast to their MSSM-like counterparts), with uncertainties at
1L~order falling to the percent~level. Of course,
$SU(2)_{\mathrm{L}}$-violating modes with a large remaining
uncertainty at the full 1L~order are also present; the exact precision
will depend on the details of the scenario. We stressed the importance
of IR~effects in the 1L~corrections (reaching~$\mathcal{O}(10\%)$ in
the considered example), underlining also that these (virtual)
corrections do not necessarily preserve the
$SU(2)_{\mathrm{L}}$~symmetry independently from the corresponding
tree-level three-body decays (radiation). The latter contribute at the
same order as 1L~two-body~decays and emerge as an essential ingredient
for a reliable determination of the widths and branching ratios at the
\mbox{complete 1L~order}.

With the completion of the evaluation of Higgs decays into
SM~final~states at 1L~order, it is possible for us to consistently
predict the total widths and branching ratios of Higgs bosons in
scenarios where the SUSY~spectrum decouples. In the converse case,
further competition can be expected from heavy Higgs decays into
\EG~electroweakinos, sleptons or even squarks. As such, our results
can also be generalized to other models based on a THDM~structure.


\appendix

\section{Bosonic decays of a heavy charged Higgs\label{ap:Hpbosdec}}

As dictated by the $SU(2)_{\mathrm{L}}$~symmetry, the decay channels
of the charged Higgs in the high-mass regime show little novelty as
compared to the case of the neutral heavy-doublet states. From the
perspective of electromagnetism, a $W$~boson is always present among
the bosonic final states (as this is the only lighter fundamental
charged boson).
		
\begin{figure}[b!]
  \centering
  \includegraphics[width=\linewidth]{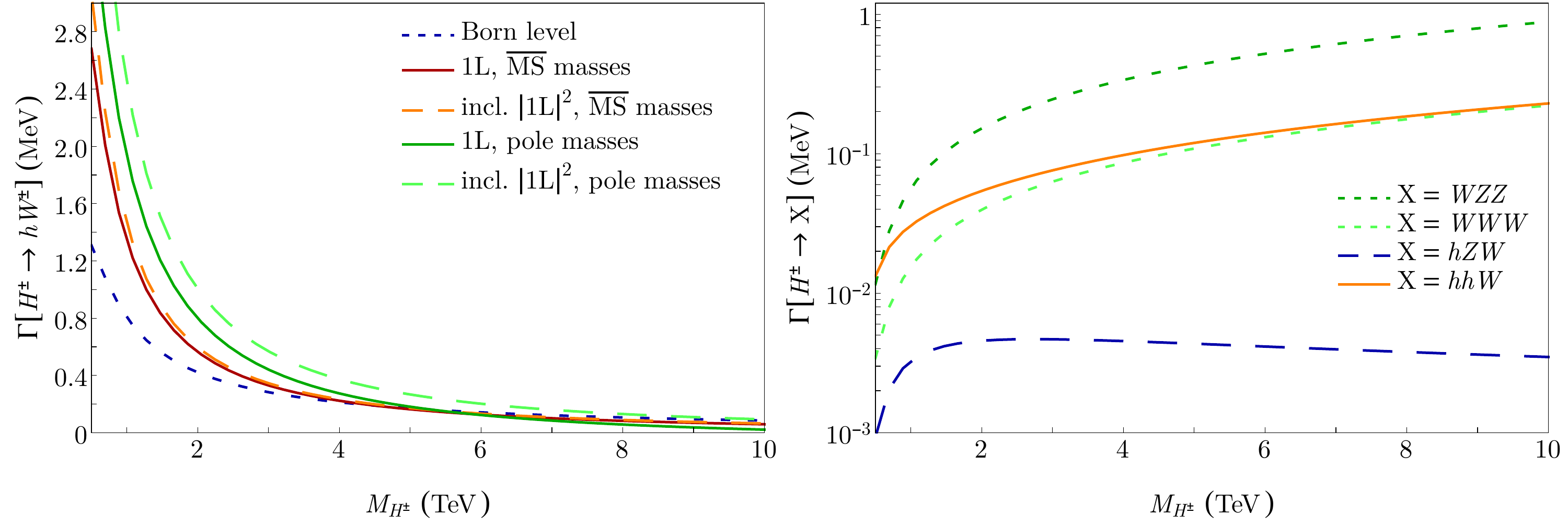}
  \caption{Bosonic Higgs decays of the charged Higgs in the MSSM
    scenario of \sect{sec:MSSM}.
    \newline{\em Left}: the two-body width $\Gamma[H^{\pm}\to
      hW^{\pm}]$ is shown at tree level (dashed blue) and at~1L for
    two choices of schemes for the Yukawa couplings (pole in green,
    vs.~\MS in orange/red), with (long-dash) or without (solid)
    1L$^2$~terms.
    \newline{\em Right}: three-body bosonic widths (non-resonant
    contributions). \label{fig:HpMSSM}}
  \vspace{3ex}
  \capstart
  \includegraphics[width=\linewidth]{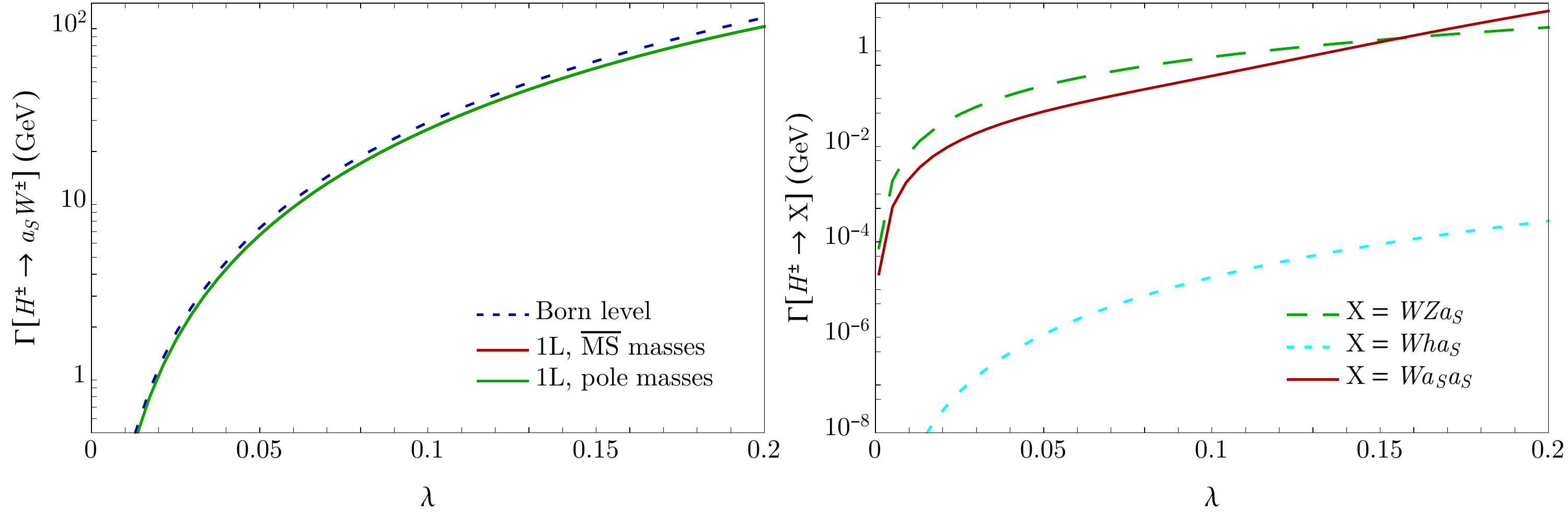}
  \caption{Bosonic Higgs decays of the charged Higgs in the NMSSM
    scenario of Sect.~\ref{sec:NMSSM}.
    \newline{\em Left}: the two-body width $\Gamma[H^{\pm}\to
      a_SW^{\pm}]$ is shown at tree-level (dashed blue) and at 1L for
    two choices of schemes for the Yukawa couplings (pole in green,
    vs.~\MS in red).
    \newline{\em Right}: three-body bosonic widths involving $a_S$
    (non-resonant contributions). \label{fig:HpNMSSM}}
\end{figure}

In the MSSM, the only kinematically accessible bosonic two-body decay
at the tree~level is~\mbox{$H^{\pm}\to h W^{\pm}$}. The
channels~$H^{\pm}\to Z W^{\pm}$ and~$H^{\pm}\to \gamma W^{\pm}$ are
purely radiative. In fact, the decay~\mbox{$H^{\pm}\to h W^{\pm}$} is
$SU(2)_{\mathrm{L}}$-violating, so that it would also vanish at the
tree~level in the exact alignment limit. Thus, similarly to the
channel~$A\to Zh$, the associated width scales
like~\mbox{$\Gamma[H^{\pm}\to h W^{\pm}]\propto
  M^2_{\text{EW}}\big/M_{H^{\pm}}$}, with the
$M^2_{\text{EW}}\big/M_{H^{\pm}}^2$~suppression emerging from the
mixing in the \CP-even sector. As we noted in \sect{sec:MSSM}, a
considerable uncertainty persists at the 1L~order on such tree-level
suppressed channels: this is illustrated by the dispersion of the
corresponding predictions in the left plot of \fig{fig:HpMSSM}, where
the choice of scheme for the Yukawa couplings or the inclusion of a
1L$^2$~term result in variations of the order of~$100\%$. The
non-resonant three-body decays in the limit~$M_{H^{\pm}}\gg
M_{\text{EW}}$ are determined by the quartic Higgs couplings and do
not necessarily break the $SU(2)_{\mathrm{L}}$~symmetry. Consequently,
they are competitive with the two-body bosonic widths as soon as
masses in the few~TeV~range are considered. The relevant widths are
shown in the plot on the right-hand side of \fig{fig:HpMSSM}.
	
In the NMSSM, the accessibility of comparatively light singlet scalar
states modifies the situation with respect to the
$SU(2)_{\mathrm{L}}$~symmetry, as unsuppressed channels
correspondingly emerge. The latter can completely dominate the charged
Higgs width if the singlet--doublet interactions are sufficiently
intense. In this case, the tree-level contribution typically
determines the widths and a percent-level precision can be expected at
the 1L~order. The width of the leading two-body channel~$H^{\pm}\to
W^{\pm}a_S$ in the scenario of \sect{sec:NMSSM} is shown on the
left-hand side of \fig{fig:HpNMSSM}, with 1L~corrections of
order~$10\%$. Once again, these 1L~contributions (for~$M_{H^{\pm}}\gg
M_{\text{EW},a_S}$) involve sizable effects of IR-type, so that
(non-resonant) three-body widths of comparable magnitude
($\simord10\%$~of the two-body width) balance them at the level of the
full width. These three-body channels involving~$a_S$ in the scenario
of \sect{sec:NMSSM} are shown on the right-hand side of
\fig{fig:HpNMSSM}. In this case, at large~$\lambda=\kappa$, they may
compete in magnitude with the MSSM-like channels.

\section{Input parameters\label{ap:input}}

The input parameters for the scenarios in \sect{sec:MSSM} and
\sect{sec:NMSSM} are summarized in \tab{tab:input}. The
SUSY~scale~$m_{\mathrm{SUSY}}$ sets the bilinear SUSY-breaking
parameters in the sfermion and electroweakino sectors; the gluino-mass
parameter is given by~$m_{\tilde{g}}$. The tree-level value
of~$A_\kappa$ is fixed according to the on-shell renormalization
scheme for the pseudoscalar singlet~$a_s$ described in
\citere{Domingo:2021jdg}; at~$\lambda=\kappa=0$ it is set to~$-3$\,GeV
such that~$M_{a_s}\sim m_{a_s}\sim300$\,GeV. At that point, it also
agrees with the input value of~$A_\kappa$ in the
$\overline{\text{DR}}$~scheme at the renormalization
scale~$\mu_{\mathrm{ren}}=m_t=172.3$\,GeV.

\begin{table}[h!]
  \centering
  \caption{Input parameters for the studied scenarios\label{tab:input}}
  \begin{tabular}{>{$}c<{$}|>{$}r<{$}@{\,}l>{$}r<{$}@{\,}l}
    &\multicolumn{2}{c}{MSSM (\sect{sec:MSSM})}
    &\multicolumn{2}{c}{NMSSM (\sect{sec:NMSSM})}\\\hline
    t_\beta & 10 && 10 &\\
    M_{H^\pm} & [0.5,10] & TeV & 3 & TeV\\
    \mu & 10 & TeV & 10 & TeV\\
    m_{\mathrm{SUSY}} & 10 & TeV & 10 & TeV\\
    m_{\tilde{g}} & 10.9 & TeV & 10.9 & TeV\\
    A_t & -2.5 & TeV & -2.5 & TeV\\
    A_{\tau,b} & 1 & TeV & 1 & TeV\\
    \lambda=\kappa & \multicolumn{2}{c}{---} & [0.01,0.20] &\\
    A_\kappa & \multicolumn{2}{c}{---} & -[3.0,7.1] & GeV
  \end{tabular}
\end{table}

\needspace{20ex}
\bibliographystyle{h-physrev}
\bibliography{literature2}

\end{document}